\documentstyle[aps]{revtex}
 
\begin{document}
 
\def\B{\mbox{\bf B}}
\def\x{\mbox{\bf x}}
\def\z{\mbox{\bf z}}
\def\u{\mbox{\bf u}}
\def\j{\mbox{\bf j}}
\def\A{\mbox{\bf A}}
\def\e{\bf {\hat e}}
\def\E{\bf {\hat E}}
\def\m{\bf {\hat m}}
\def\M{\bf {\hat M}}
\def\s{\bf {\hat s}}
\def\S{\bf {\hat S}}
\def\v{\bf v}
\def\zhat{\hat {\bf z}}
\def\psibar{\hat \psi}
\def\V{\mbox{\bf V}}
\def\u{\mbox{\bf u}}

\draft

\title{Advection and diffusion in a three dimensional chaotic flow}
\author{X. Z. Tang\footnote{Email: tang@chaos.ap.columbia.edu}}
\address{Department of Applied Physics, 
Columbia University, New York, NY 10027}
\author{A. H. Boozer}
\address{Department of Applied Physics, Columbia University, New York,
NY 10027\\
and Max-Planck Institut f{\"u}r Plasmaphysik, Garching, Germany}
\date{\today}

\maketitle

\begin{abstract}

The advection-diffusion equation is studied via a global Lagrangian
coordinate transformation. The metric tensor of the Lagrangian coordinates
couples the dynamical system theory rigorously into the solution of this
class of partial differential equations. 
If the flow has chaotic streamlines, the diffusion will dominate the 
solution at a critical time, which scales logarithmically with the 
diffusivity. The subsequent rapid diffusive relaxation is completed
on the order of a few Lyapunov times, 
and it becomes more anisotropic the smaller the diffusivity. 
The local Lyapunov time of 
the flow is the inverse of the finite time Lyapunov exponent.
A finite time Lyapunov exponent can be expressed in terms of two
convergence functions which
are responsible for the spatio-temporal complexity of both the advective
and diffusive transports.
This complexity gives a new class of diffusion barrier in the
chaotic region and a fractal-like behavior in both space and time.
In an integrable flow with shear, there also exist fast and slow diffusion.
But unlike that in a chaotic flow, a large gradient of the scalar field
across the KAM surfaces can be maintained since the fast diffusion 
in an integrable flow
is strictly confined within the KAM surfaces.   
 
\end{abstract}

\pacs{PACS numbers: 47.10.+g, 52.30.-q, 05.45.+b\\
\\
Keywords: Advection-diffusion equation, finite time Lyapunov exponent,
	${\s}$ line, diffusion barriers, hamiltonian chaos, fractal, mixing}

\section{Introduction}
\label{section:intro}

\subsection{Motivation}
\label{subsection:moti}

The transport of a passive scalar $\phi$ embedded in a fluid flow is 
governed by the advection-diffusion equation \cite{landau},
\begin{equation}
\label{s1e1}
\partial\phi/{\partial t}+{\v}\cdot\nabla\phi=-(\nabla\cdot\Gamma_d)/\rho
\end{equation}
with ${\v}({\x},t)$ the fluid velocity, $\rho$ the fluid density, and
$\Gamma_d = - \rho D \nabla\phi$ the diffusive flux.
In its most primitive form, $D$ is just the molecular diffusivity which
is typically a small number, giving rise to an unphysically long
characteristic diffusive time scale $L^2/D.$ Being a linear equation, the 
non-triviality of the advection-diffusion equation comes from the flow
velocity field ${\v}({\x},t).$ 
The purpose of this paper is to illustrate the general properties of
the solution to the advection-diffusion equation in the case of a
three dimensional chaotic flow.
Our method \cite{tang} is based on a global Lagrangian coordinate
transformation, which rigorously couples the dynamical system
theory \cite{ruelle85} into the solution of the advection-diffusion equation.
The finite time Lyapunov exponent and the geometry of the so-called
${\s}$ lines play the central roles in our theory\cite{tang}, which is 
a distinct feature from other works in this area. 
 
The standard treatment \cite{taylor,zeldovich88a} of the advection-diffusion 
equation presumes 
a turbulent background flow, the so-called turbulent mixing problem.
A method of averaging [e.g. \cite{zeldovich88a}] is employed to separate the 
rapidly fluctuating component from the statistical mean.
The effect of the fluctuating flow component is then modeled as an
effective diffusivity \cite{taylor} leading to enhanced mixing. 
This approach is justified by the wide separation of 
the correlation scales in the mean and the fluctuating flow velocity 
components \cite{taylor,zeldovich88a,krommes}.

Actually 
a smooth, non-turbulent, flow with chaotic fluid trajectories
gives fundamentally different solutions 
to the advection-diffusion equation than a non-chaotic,
or integrable, flow.
The smoothness of the flow field precludes the method of averaging employed
in the turbulent mixing theory.
Existing literature on chaotic mixing largely concerns with the ideal 
advection equation [$D=0$ in equation (\ref{s1e1})] whose solution 
is found by following the Lagrangian trajectories, for more details see
section \ref{subsection:review}.
The diffusive effect, i.e. right-hand-side of equation (\ref{s1e1}),
is usually treated as an add-on on the Lagrangian trajectory picture.
For example, it has been modeled by a stochastic perturbation to the Lagrangian
trajectories [{\it i.e.} a Langevin equation], or by a Gaussian smoothing kernel
on the Lagrangian trajectories. 

Most previous work [see section \ref{subsection:review} for detail] 
on chaotic mixing 
circumvents the advection-diffusion equation by working with
the Lagrangian description of the fluid directly.
This permits the utilization of dynamical system theory, particularly the
multiplicative ergodic theorem\cite{oseledec} and the geometrical method,
which leads to new insights unavailable in the usual Eulerian picture.
A numerical solution of
equation (\ref{s1e1}) using an Eulerian 
PDE solver has no obvious connection with the crucial features like KAM islands,
 chaotic components,
or Lyapunov exponents.
On the other hand, the efficiency and reliability of numerical PDE solvers for 
following the global solution for a long time 
requires an understanding of the
properties of the solution and the features that define 
the range of validity of the numerical scheme. 
This is the primary motivation for our previous work in the two dimensional case 
and the current treatment for the three dimensional case.  
  
The basic difference between our approach and those reviewed 
in section \ref{subsection:review} is
that we directly solve the
advection-diffusion equation including the effects of a finite diffusivity.
The need to keep the diffusion term, despite the smallness
of $D,$ will become obvious once we obtain the full solution.
The most obvious requirement for keeping a small $D$ comes from the
time for diffusion [the right-hand side of 
equation (\ref{s1e1})] to dominate the solution,
which has a logarithmic dependence on $D.$  
In a general context, the diffusion term is a singular perturbation to the 
pure advection equation.
Mathematically it
changes the characteristics, or type, of the underlying PDE. Physically 
it is responsible for removing the time reversibility of 
the physical process. 
This can be quantitatively explained by examining the mean variance of the
passive scalar $\phi,$
$$
S \equiv -\int (\phi^2/2) d^3{\x}.
$$
For a bounded system, it is straightforward to show that the entropy-like quantity
$S$ would increase or saturate only if $D$ does not vanish, see equation (\ref{s1e7}).  

In the next section we will briefly review some related work on chaotic mixing,
ranging from chaotic advection, hamiltonian transport theory,
to fractional kinetic theory.   
Neither the list of topics nor the literature cited can be exhaustive, but they
give a reasonable perspective for contrasting our approach and
results. For those familiar with the literature, section \ref{subsection:review}
can be skipped in its entirety.

\subsection{A brief review of related work}
\label{subsection:review}

As one of the primary physical applications of the chaos theory,
it was recognized \cite{aref} in the eighties that 
smooth (laminar) flow could also lead to efficient
mixing, as long as the flow trajectories are non-integrable or chaotic
[for a sampling of experimental work, see \cite{experiment1,experiment2}].
Despite the variants in name (chaotic advection or 
Lagrangian turbulence\cite{aref},
hamiltonian transport theory\cite{meiss92}, 
and fractional kinetic theory\cite{zaslavsky}),
these treatments make the same assumption of ignoring the 
right-hand-side (diffusion) term in equation (\ref{s1e1}) and are,
therefore, concerned with an ideal advection equation:
\begin{equation}
\label{advection_equation}
\partial\phi/\partial t + {\v}({\x},t)\cdot\nabla\phi=0, \,\,\,\,
{\rm or} \,\,\,\, d\phi/dt = 0.
\end{equation}
The mathematical solution to the advection equation is 
remarkably straightforward.
Since the passive scalar is frozen into the fluid element, 
the distribution function $\phi$
at arbitrary time is found by following the trajectory of 
each fluid element,
\begin{equation}
\label{trajectory}
d{\x}({\xi},t)/dt = {\v}({\x},t),
\end{equation}
with the initial condition ${\x}({\xi},t=0) = \xi.$
After integrating  equation (\ref{trajectory})
to obtain ${\x}({\xi},t)$
the solution to the ideal advection equation is 
 $\phi({\x}({\xi},t),t) = \phi({\xi},t=0).$ 

Reducing the solution of the advection equation to an
integration of flow trajectories not only simplifies the problem,
but also allows
the theory of dynamical systems to be utilized for
classifying the trajectories and the associated
mixing properties\cite{ottino}. 
The crudest estimate is entirely based on topology and the intuitive 
criteria that ergodicity implies good mixing.
A plot of the Poincare section of the flow then becomes the standard tool. 
For example, on the Poincare section of a two-dimensional
time-periodic flow, integrable trajectories lie on topological circles
(so-called regular component, or KAM tori in hamiltonian mechanics)
but non-integrable or chaotic trajectories fill a finite area (so-called 
irregular component). 
Integrable trajectories can divide the space into an infinite set of
ergodic subregions.
The integrable or regular, regions consist of  
closed lines or surfaces, so they are poor for mixing.
Only chaotic regions, {\it i.e.}, 
the irregular components, occupy a finite volume,
which is required for good mixing.
It is still an unresolved mathematical problem
whether the irregular component occupies a 
finite measure. The difficulty lies 
in the fact that numerous regular components are embedded in an irregular 
component and the summation of those infinitely many small regular regions
may not be small. 
However, this subtlety is not crucial for physical applications.
If one goes back to the original advection-diffusion equation, any small but 
finite diffusivity would impose a cut-off for the 
smallest spatial scale on which one needs
to worry about the regular islands. This effectively guarantees that a 
chaotic zone occupies finite volume for the purpose of 
passive scalar transport.  

Although chaotic advection and hamiltonian transport theory in principle
are entirely different subjects, they are often mathematically
equivalent. An example is a divergence-free, two-dimensional time-periodic
fluid flow, which is mathematically equivalent to
 a one and a half degree of freedom hamiltonian system.
A major advance in understanding was the discovery of the cantorus\cite{perceval}, 
an invariant curve or surface
similar to an KAM surface but with the crucial difference of 
its being on a cantor 
set. In other words, a cantorus is a KAM-like structure but with 
numerous holes so it can
not separate an irregular component.
The trajectories `leak' through the cantori in a peculiarly orderly fashion, 
much like going through a revolving door, 
hence the name turnstile\cite{meiss_turnstile}.  
The cantori act as practical borders partitioning 
one ergodic irregular component
into different subcomponents which have fast mixing within but much slower
advective transport across. Motivated by this separation of time scales, 
a Markov tree model\cite{meiss_markov} was 
introduced to describe the slow mixing between the 
subcomponents separated by the 
cantori. Another less efficient but mathematically 
rigorous approach, the so-called lobe dynamics\cite{wiggins_lobe}, 
generalized the idea 
of turnstiles for the advective transport across the boundary set by the
invariant manifolds of any hyperbolic or normally hyperbolic sets.
In effect, it calculates the advective flux by following the trajectories
through a sequence of ``revolving doors'' (turnstiles) on the 
otherwise closed boundaries.
The boundaries are formed by the global invariant manifolds of  hyperbolic
sets, so they in principle could provide an arbitrarily 
fine partition of the space. 
In practice, the lobe dynamics become exponentially 
complicated as time proceeds
so it is usually thought to be applicable for only a short time.
The same argument also shows it is suitable for tracing initially isolated 
distribution but not for following the time evolution of a 
spatially-extended initial distribution.  
The reliance on the invariant manifolds also limits the application since
their existence is not obvious for time quasi-periodic or aperiodic flows.

In the same spirit as using a 
diffusion equation to model Brownian motion\cite{einstein},
statistical approaches have also been advanced to model 
the chaotic advective transport as a diffusion process alone. 
The earliest work of this kind was due to 
Rechester and Rosenbluth\cite{rechester}
who modeled the chaotic motion of the electrons due to stochastic magnetic
field lines in a toroidal magnetic
confinement device by an effective diffusion coefficient. 
The Rechester-Rosenbluth diffusivity is unusual since it is 
based on a quasi-linear calculation\cite{rosenbluth} 
for the rate of average squared separation. 
If following the standard definition\cite{einstein} 
one finds such a diffusion
coefficient could diverge for deterministic chaotic trajectories, the 
so-called super- or hyper-diffusion phenomena\cite{superdiffusion}.
One elegant solution is due to Zaslavsky \cite{zaslavsky} who recognized
that the island-chain structure in the phase
space of a hamiltonian system (also called stochastic layer)
has a fractal dimension in both space and time. Modifying the definition
of diffusivity using fractional powers, one can obtain a convergent 
 diffusivity. The resulting
Fokker-Planck equation has fractional derivatives 
in both time and spatial
coordinates, hence the name fractional kinetic equation.    
Novel as it is conceptionally , this approach has not matured to the stage 
of demonstrating a practical application.
Its applicability to highly chaotic flows is also not clear since the surviving
islands are usually of negligible size.

\subsection{Outline of our approach and results}
\label{subsection:results}

The Lagrangian nature of the chaotic advection and the desire for a solution
to advection-diffusion equation in Eulerian frame imply the need for a method
that relates the two in the presence of a small but finite diffusivity.
A straightforward approach would be to solve the advection-diffusion equation
in Lagrangian coordinates, the coordinate system directly associated with
the Lagrangian description of fluids. 
Although the Lagrangian description of a fluid flow is widely known,
global Lagrangian 
coordinates were not applied to the advection-diffusion problem until recently \cite{tang}.
The validity of this approach is based on the general principle
that the description of a physical phenomena is independent of the choice
of the coordinate system. The solution in Lagrangian coordinates
is non-trivial due to the metric tensor, which arises in the 
$\nabla^2$ operator of the diffusion term and is highly anisotropic
if the flow is chaotic. 
The metric tensor rigorously
couples the solution of the advection-diffusion equation to the dynamical system theory.
It is also the metric tensor that bridges the Lagrangian picture and the
Eulerian solution.

One unique feature of the use of global Lagrangian coordinates
 is the ability to treat flow fields that are
far from integrable and thus highly chaotic, an area where less is known
but which is of great importance \cite{ottino}.
Although the integrable case is rigorously 
treated in section \ref{section:integrable},
we will center our discussion on highly chaotic flows
because of their practical importance. 
By definition, a flow is chaotic if the distance $\delta$ between neighboring 
fluid elements tends to vary (diverge or converge) exponentially in time, 
$\delta\propto\delta_0\exp(\lambda t),$ with $\lambda$ the Lyapunov exponent. 
In advection-diffusion problems, the Lyapunov time $1/\lambda$ associated
with the most negative Lyapunov exponent 
defines a natural characteristic
time scale for a chaotic flow. The characteristic diffusion time $L^2/D$
is determined by the diffusivity $D$ and the typical spatial scale $L$ 
of the initial gradient of $\phi.$
Our previous analysis \cite{tang} 
illustrated that the characteristic dimensionless parameter of the chaotic 
transport problem in two dimensions is the ratio of the 
characteristic diffusion time and the 
Lyapunov time of the flow, {\it i.e.} $\Omega\equiv \lambda L^2/D.$
The diffusivity is generally very small so $\Omega\gg 1.$
For $\Omega\gg 1$ the passive scalar undergoes a pure advection period until
the time $t_a\equiv\ln(2\Omega)/2\lambda.$ A rapid diffusive relaxation
removes the spatial gradient of the passive scalar during a period of a few
Lyapunov time $1/\lambda$ centered on the time $t_a.$ 
This diffusion is of one
dimensional since it only occurs along the ${\s}_\infty$ direction, 
which defines the stable direction for neighboring streamlines to converge.
In generic flows, the finite time Lyapunov exponent is a function of both
time and position. It was found for 2D systems that 
the geometry of the field 
line of the ${\s}_\infty$ vector determines the 
spatial variation of the finite time
Lyapunov exponent along the ${\s}$ lines, and hence the local 
diffusive transport \cite{tang}. Diffusion is impeded at the sharp bends of 
an ${\s}$ line, which has a peculiarly small finite time Lyapunov exponent.

Mixing in a three
dimensional flow is clearly important for practical applications.
The goal of this paper is to carry out an explicit analysis in 3D and 
establish a similar level of understanding as 
previously achieved in 2D \cite{tang}. 
As we will show, the main physics results obtained in 2D apply
to 3D situations. These include the characteristic time scales for advection
and diffusion which are determined by the Lyapunov time and the dimensionless 
number $\Omega,$ and the extreme anisotropy of the diffusive relaxation. 
Of course, the increase of spatial dimension from two to three does introduce
new subtleties in the solution, 
some of which will be discussed in the main text.

The main body of the paper is organized as follows. 
Section \ref{section:overview} gives a description of the problem
and an overview of our results, which is the minimum amount of material 
necessary for understanding the thesis of this paper. 
In section \ref{section:coordinates}, the advection-diffusion equation is
solved for a three dimensional flow in natural Lagrangian coordinates.
The properties of the finite time Lyapunov exponent and diffusion barriers
are discussed in section \ref{section:exponent}.
In section \ref{section:integrable} transport in an integrable region of
a three dimensional flow is treated.
Some models of chaotic flows and numerical illustrations are given in
section \ref{section:flow}.

\section{Overview}
\label{section:overview}

Chaos and its effect on diffusive transport in a fluid flow 
can be conveniently examined
using Lagrangian coordinates \cite{comment_lagrangian}. 
The motion of a fluid element is described by the differential equation
\begin{equation}
\label{s1e2}
d{\x}/{dt} = {\v} ({\x}, t).
\end{equation}
The trajectory in real space is the solution
${\x} = {\x}(\xi, t)$ with $\xi = {\x}(\xi, t = 0)$ the Lagrangian coordinates.
The distance between neighboring fluid points at time $t$ is related to their
initial separation by $d{\x}\cdot d{\x} = g_{ij} d\xi^i d\xi^j,$ 
where $g_{ij} \equiv 
(\partial {\x}/\partial \xi^i)\cdot (\partial {\x}/\partial \xi^j)$ is the 
metric tensor of the Lagrangian coordinates. The matrix inverse
of $g_{ij}$ is
$g^{ij} \equiv \nabla\xi^i\cdot\nabla\xi^j.$ The Jacobian of the Lagrangian 
coordinates is related with the metric tensor 
by $J^2 = \|g_{ij}\| = 1/\|g^{ij}\|.$ For a divergence-free flow $J=1.$
The metric tensor is a positive definite, symmetric matrix, 
so it can be diagonalized 
with real eigenvectors and positive eigenvalues, {\it i.e.},
$$
g_{ij} = \Lambda_l {\e}{\e} + \Lambda_m {\m}{\m} + \Lambda_s {\s}{\s}
$$ 
with the three positive eigenvalues $\Lambda_l\ge\Lambda_m\ge\Lambda_s>0.$  
The Lyapunov characteristic exponents of the flow are given by
\begin{eqnarray}
\lim_{t\rightarrow\infty} \ln\Lambda_l/2t & = & \lambda_l^\infty, \nonumber\\
\lim_{t\rightarrow\infty} \ln\Lambda_m/2t & = & \lambda_m^\infty, \nonumber \\
\lim_{t\rightarrow\infty} \ln\Lambda_s/2t & = & \lambda_s^\infty,\nonumber 
\end{eqnarray}
and the eigenvectors ${\e},{\m}$ and ${\s}$ have 
well defined time asymptotic limits:
${\e}_\infty, {\m}_\infty$ and ${\s}_\infty$ 
(see appendix \ref{appendix:ergodic}).
In those regions where there is a positive Lyapunov exponent, the
flow is said to be chaotic, otherwise it is said to be integrable.

For a three dimensional divergence-free flow 
with symmetry under time reversal, 
the set of Lyapunov characteristic exponents is 
symmetric with respect to zero, {\it i.e.},
$\lambda_l^\infty=-\lambda_s^\infty=\lambda^\infty > 0$ and
$\lambda_m^\infty = 0.$
The time asymptotic eigenvectors determine the asymptotic behavior of
neighboring fluid elements. Along ${\s}_{\infty}$ (${\e}_{\infty}$) direction,
neighboring points converge (diverge) exponentially in time. 
But their separation varies
only algebraically with time along ${\m}_\infty$ direction. 
We note that ${\s}_\infty$ defines the stable direction in a chaotic
flow (see appendix \ref{appendix:ergodic}). 
The ${\s}_\infty$ is a smooth function of
position so it gives rise to a vector field. The field lines of the
${\s}_\infty$ vector are called ${\s}$ lines. 
It should be pointed out that  time-dependent flows
might not always have the middle Lyapunov 
exponent zero. Although our analysis is presented in the case of $\lambda_m=0$
for the sake of clarity, the more complicated case with an arbitrary
combination of positive and negative Lyapunov exponents can also be treated.
In fact, a clear understanding of the simple case makes 
the physics of the more complicated
case transparent, for details, see appendix \ref{appendix:3D}. 

Although the infinite time Lyapunov exponents are better known
in mathematics, their finite time analogies are of greater interests 
in physics.
The finite time Lyapunov exponents,
\begin{equation}
\label{finite_time}
\lambda_l(\xi,t)\equiv(\ln\Lambda_l)/2t;\,\,\,
\lambda_m(\xi,t)\equiv(\ln\Lambda_m)/2t;\,\,\,
\lambda_s(\xi,t)\equiv (\ln\Lambda_s)/2t,
\end{equation}
are functions of position $\xi$ and time.
We find that the finite time Lyapunov exponent $\lambda_s$ and
its associated ${\s}_\infty$ vector play the most important role
in diffusive transport.
This can be seen by transforming the advection-diffusion equation,
Eq. (\ref{s1e1}), into Lagrangian coordinates.
For the simplicity of notation, we define
\begin{equation}
\label{lyapunov_exponent}
\lambda(\xi,t) \equiv - \lambda_s(\xi,t).
\end{equation}
The inverse of $\lambda,\,1/\lambda,$ is the Lyapunov time of the flow. 
We wish to point out again that the most negative Lyapunov exponent ($\lambda$)
defines the characteristic Lyapunov time of the flow for 
advection-diffusion problems.  

In Lagrangian coordinates, the advection-diffusion equation becomes an ordinary
diffusion equation with a tensor diffusivity $D^{ij}= D g^{ij}$ \cite{tang},
\begin{equation}
\label{s1e3}
\Bigl({\partial\phi\over{\partial t}}\Bigr)_{\xi} = {1\over\rho_0}
 \sum {\partial\over{\partial
{\xi}^i}} \rho_0 D^{ij} {\partial\phi\over{\partial{\xi}^j}}
\end{equation}
where $\rho_0(\xi)$ is the initial fluid density profile,
$\rho_0({\xi}) = \rho({\xi}, t=0).$ The magnitude of the gradient of $\phi$ is
 given by 
\begin{equation}
\label{s1e4}
(\nabla\phi)^2 = \sum (\partial\phi/\partial{\xi}^i) g^{ij} (\partial
\phi/\partial{\xi}^j).
\end{equation}
The effect of the flow on the evolution of a passive scalar is, therefore, 
determined by the metric tensor of Lagrangian coordinates $g^{ij}.$
For simplicity we will assume that the initial fluid density distribution
 $\rho_0(\xi)$ is a constant. The diffusion equation, Eq. (\ref{s1e3}), can
 then be written in Lagrangian coordinates as
\begin{equation}
\label{s1e5}
\partial\phi/\partial t = - \nabla_0\cdot\gamma \hspace {1.0cm}
{\rm with} \hspace{0.5cm} \gamma \equiv - {\tensor{\bf D}}\cdot\nabla_0\phi
\end{equation}
and ${\tensor{\bf D}}$ the tensor diffusivity $D^{ij}.$ 
Here $\nabla_0$ denotes gradient in Lagrangian coordinates.
Equation (\ref{s1e5}) maximizes
an entropy-like quantity
\begin{equation}
\label{s1e6}
S\equiv - \int (\phi^2/2) d^3\xi
\end{equation}
while holding $\int\phi d^3\xi$ constant. The time derivative of $S$ is
\begin{equation}
\label{s1e7}
dS/dt = - \int(\gamma\cdot\nabla_0\phi)d^3\xi.
\end{equation}
So the entropy production rate per unit volume is positive definite and
given by
\begin{equation}
\label{s1e8}
\dot{s}(\xi,t) \equiv - \gamma\cdot\nabla_0\phi.
\end{equation}
Only diffusion creates entropy and removes the time reversibility
of the system. Even a tiny diffusivity $D$ leads to an inevitable rapid
entropy production in a chaotic flow. To see that, let's substitute
$g^{ij}={\e}{\e}/\Lambda_l + {\m}{\m}/\Lambda_m + {\s}{\s}/\Lambda_s$ into
equation (\ref{s1e8}),
\begin{equation}
\label{s1e9}
\dot{s} = D ({\e}\cdot\nabla_0\phi)^2 e^{-2\lambda_l t}
+ D ({\m}\cdot\nabla_0\phi)^2 e^{-2\lambda_m t}
+ D ({\s}\cdot\nabla_0\phi)^2 e^{2\lambda t}.
\end{equation}
Since $\lambda_s < 0$ and $\lambda\equiv-\lambda_s>0,\, \dot{s}$ 
would grow exponentially in time without bound unless
the diffusion intervenes and quickly removes the coordinate dependence 
of $\phi$ along the ${\s}$ lines. This result holds independent of
the smallness of $D,$  as long as it does not vanish.
Such a conclusion can also be obtained by examining the tensor
diffusivity.
The effective diffusivity along the ${\e}_\infty$ direction is
negligible since
$$
D_{ee} \equiv {\e}_\infty\cdot\tensor{\bf D}\cdot{\e}_\infty
\approx D / \exp (2\lambda_l t)
$$
with $\lambda_l > 0.$
The effective diffusivity along the ${\m}_\infty$ direction 
$$
D_{mm} \equiv {\m}_\infty\cdot\tensor{\bf D}\cdot{\m}_\infty
\approx D / \Lambda_m
$$
is small if $D$ is small.
In contrast, the effective diffusivity along the ${\s}_\infty$
direction grows exponentially in time,
$$
D_{ss} \equiv {\s}_\infty\cdot\tensor{\bf D}\cdot{\s}_\infty
\approx D / \Lambda_s \approx D \exp(2\lambda t).
$$
The exponential amplification of the effective diffusivity along
the ${\s}_\infty$ direction in Lagrangian coordinates corresponds
to an exponentially growing gradient of the passive scalar in real
space. It is easy to see that the diffusion becomes a dominant
process in a chaotic flow for $\lambda t\gg 1$ regardless of how
small $D$ may be.
Furthermore, the diffusion, once it becomes important, is highly
anisotropic.  

The strong anisotropy of the diffusion process demands care in
the choice of coordinates. The rapid diffusion, which occurs only
along an ${\s}$ line, can be confined to one coordinate if the 
coordinate system is chosen appropriately. We named such coordinate
system natural Lagrangian coordinates \cite{tang_thesis} and showed how to
construct them in 2D in \cite{tang}. In this paper, we give a form of
natural Lagrangian coordinates in three dimensional space, which then 
allows us to obtain the 
general properties of the solution to the advection-diffusion equation
in a three dimensional chaotic flow. The findings agree with our earlier
results in 2D \cite{tang}.
In summary, the characteristic dimensionless parameter $\Omega$
for the chaotic transport of a passive scalar is the ratio of the 
characteristic diffusion time and the Lyapunov time of the flow,
\begin{equation}
\label{s1e10}
\Omega\equiv \lambda L^2/D,
\end{equation}
with $L$ the typical spatial scale and $\lambda$ the Lyapunov exponent,
equation (\ref{lyapunov_exponent}).
If the characteristic diffusion time $L^2/D$ is much longer than the 
Lyapunov time $1/\lambda$ of the flow, {\it i.e.}, $\Omega\gg 1,$ 
the chaotic transport is given by ideal advection (the scalar is 
carried by the fluid element along its trajectory) for time less than
$t_a - 1/2\lambda$ with $t_a\equiv (\ln2\Omega)/2\lambda.$
The ideal advection causes the gradient of the scalar field to increase
by a factor of $\Omega.$ Then a rapid diffusion occurs and causes the 
flattening of the gradient and associated entropy production during
a relatively short interval $1/\lambda$ centered on $t_a.$ This rapid
diffusive relaxation occurs only along the ${\s}$ lines, which is a 
special feature for chaotic flows.

The existence of a characteristic chaotic transport time scale
$t_a$ implies that the finite time Lyapunov exponent rather than the
infinite time Lyapunov exponent determines the chaotic transport.
The spatio-temporal complexity of the diffusive transport,  
as reflected in the entropy production rate per unit volume $\dot{s},$
is determined by the finite time Lyapunov exponent $\lambda(\xi,t).$
 For example,
the places with significantly smaller $\lambda(\xi,t)$ (hence $\Omega$)
pose barriers for the diffusive transport and entropy production.
It must be emphasized that once the flow field is specified, the 
finite time Lyapunov exponent $\lambda(\xi,t)$ is completely determined.

Numerical results suggest that the finite time Lyapunov exponent 
$\lambda(\xi,t)$ of a three dimensional conservative system can be
decomposed into three parts,
\begin{equation}
\label{s1e11} 
\lambda(\xi,t) = \tilde{\lambda}(\xi)/t + 
f(\xi,t)/\sqrt{t} + \lambda^\infty,
\end{equation}
where
\begin{equation}
\label{s1e12}
{\s}_\infty\cdot\nabla_0 f(\xi,t) = 0
\end{equation}
and $\lambda^\infty$ is the infinite time Lyapunov exponent.
The spatial dependence of the finite time Lyapunov exponent $\lambda(\xi,t)$
is related to the geometry of the ${\s}$ line through $\tilde{\lambda}$ by
\begin{equation}
\label{s1e13}
{\s}_\infty\cdot\nabla_0\tilde{\lambda}(\xi) + \nabla_0\cdot{\s}_\infty=0,
\end{equation}
where $\tilde{\lambda}(\xi)$ is a smooth function of position due to
the smoothness of the vector field ${\s}_\infty.$
Hence we have, once again, directly related the geometry of an ${\s}$ line
to the diffusive transport through $\lambda(\xi,t).$ These new results for
three dimensional systems, equations (\ref{s1e11},\ref{s1e12},\ref{s1e13}),
have the exactly same form as what we found for two
dimensional conservative systems \cite{tang} [of course, the number of spatial
coordinates is now three in equation (\ref{s1e13})].  
Just like in 2D, the function $\tilde{\lambda}$ in equation (\ref{s1e11})
is responsible for the description of barriers for diffusive transport
in chaotic flows while the function $f(\xi,t)$ characterizes the fractal 
nature of the chaotic advection \cite{experiment1,ott} and 
the chaotic diffusive transport \cite{tang}.

Fast diffusion occurs along the ${\s}$ lines and 
entropy production rate $\dot{s}$
is given by the finite time Lyapunov exponent $\lambda(\xi,t).$
Equations (\ref{s1e11},\ref{s1e12},\ref{s1e13}) imply that $\lambda(\xi,t)$
varies little if the ${\s}$ line is straight 
(hence small $\nabla_0\cdot{\s}_\infty$). But $\lambda(\xi,t)$ will have a 
strong variation where the ${\s}$ line has a sharp bend. 
Our numerical results show that $\lambda(\xi,t)$ makes a sharp dip at
the sharp bends of the ${\s}$ lines, for an illustration, see figure 8.
Small $\lambda$ leads
to a small $\Omega.$ Hence diffusive transport is impeded on the sharp bends
of the ${\s}$ lines and a class of diffusion barriers is created inside
the chaotic region of the flow. Our results on diffusion barriers in 2D
chaotic flows, therefore, have been reestablished in three dimensional flows.

We find that there are also spatially separated fast diffusion and
slow diffusion in an integrable region
of a shear flow.
The fast diffusion is confined to the KAM surfaces only, 
while the slow diffusion occurs
across the good KAM surfaces.
It is the radial direction across the nested KAM surfaces in 
which  a significant gradient 
of the scalar field can be maintained. 
For a detailed analysis, see section \ref{section:integrable}.
 
\section{Solving the advection-diffusion equation in natural Lagrangian
coordinates}
\label{section:coordinates}

The tensor diffusivity $Dg^{ij}$ with
$g^{ij}=\Lambda_l^{-1}{\e}{\e} +\Lambda_m^{-1}{\m}{\m} + \Lambda_s^{-1}{\s}{\s}$
is strongly anisotropic due to the chaotic nature of the flow. 
By introducing a set of new Lagrangian coordinates 
in which the large component of diffusion
affects only one coordinate, we can  simplify the computation and understand
the general properties of the chaotic transport of passive scalars. A coordinate
system that has this property is called 
natural Lagrangian coordinates \cite{tang_thesis}.

In two dimensions, the metric tensor is $g^{ij}=\Lambda_l^{-1}{\e}{\e} +
\Lambda_s^{-1}{\s}{\s},$  and the curl of ${\e}_\infty$ and ${\s}_\infty$ 
will be orthogonal
to themselves. That is, 
if we write ${\e}_\infty = e_x {\hat {\bf x}} + e_y {\hat {\bf y}}$
and ${\s}_\infty=s_x {\hat {\bf x}} + s_y {\hat {\bf y}},$ the curls 
will lie along the ${\hat {\bf z}}$ axis.
 Hence  natural Lagrangian coordinates $\alpha$-$\beta$ 
can be defined by ${\e}_\infty = a\nabla\alpha$ and ${\s}_\infty = b\nabla\beta$
with Jacobian
 $J_{\alpha\beta} = ab$\cite{tang}, 
using the orthonormality of ${\e}_\infty$ and
${\s}_\infty,$ ${\s}_\infty\cdot\nabla\alpha=0$ and
 ${\e}_\infty\cdot\nabla\beta=0.$

In three dimensions, we can 
establish $\alpha$-$\beta$-$\zeta$ coordinates such that 
${\s}_\infty\cdot\nabla\alpha={\s}_\infty\cdot\nabla\zeta=0,$ 
but in general one can 
not choose the other coordinate $\beta$ 
such that ${\e}_\infty\cdot\nabla\beta = 
{\m}_\infty\cdot\nabla\beta=0.$
To separate out the large component of diffusion, 
the ${\s}_\infty$ vector must satisfy
${\s}_\infty\cdot\nabla\alpha=0$ and 
${\s}_\infty\cdot\nabla\zeta=0.$ The coordinates
$\alpha$-$\beta$-$\zeta$ given by the following equations
\begin{eqnarray}
\label{s2e1}
\nabla\alpha & = & f {\e}_\infty + g {\m}_\infty \nonumber\\
\nabla\zeta & = & p {\e}_\infty + q {\m}_\infty \\
\nabla\beta & = & a {\e}_\infty + b {\m}_\infty + c {\s}_\infty\nonumber
\end{eqnarray}
satisfy this requirement.
The functions $f, g, p, q, a, b$ and $c$ are determined locally by the 
properties of ${\e}_\infty, {\m}_\infty$ and ${\s}_\infty$
from a set of first order differential equations 
(for a proof that such a coordinate
system exists in the neighborhood of an arbitrary point, 
see appendix \ref{appendix:coordinates}).
The Jacobian of the $\alpha$-$\beta$-$\zeta$ coordinates is 
$J_n = 1/(fq - gp)c.$
In both 2D and 3D cases, the $\beta$ coordinate gives the 
direction of rapid diffusion.
Diffusion in the other coordinate(s) is either 
severally suppressed  or can not be distinguished from that in an
integrable flow. 

The infinite time Lyapunov exponent is a constant in one ergodic region.
The finite time Lyapunov exponents, as defined in equation (\ref{finite_time}),
are functions of position and time.
The eigenvectors of the metric tensor ${\e}, {\m}$ and ${\s}$ 
also depend on position
and time.  They  converge to time independent functions of Lagrangian 
position, the time asymptotic eigenvectors 
${\e}_\infty(\xi),\ {\m}_\infty(\xi)$ and ${\s}_\infty(\xi),$
for $\Lambda_s^{-1}\gg 1.$ 
The convergence of the ${\s}$ vector is of most importance, and the finite time
${\s}$ is related to the asymptotic eigenvectors by
\begin{equation}
\label{s2e2}
{\s} \propto {\s}_\infty + \sigma_m \Lambda_s {\m}_\infty 
+ \sigma_e \Lambda_s {\e}_\infty
\end{equation}
where $\sigma_m(t)$ and $\sigma_e(t)$ depend algebraically on time and 
measure the rate of convergence, for a numerical illustration see Fig. 1 
in section \ref{section:flow}.

The $\alpha$-$\beta$-$\zeta$ coordinate system with Jacobian $J_n =
1/(fq - gp)c$ simplifies the diffusion equation,
Eq. (\ref{s1e5}). In these coordinates one has
\begin{equation}
\label{s2e3}
{\partial\phi\over{\partial t}} = - {1\over J_n} {\partial\over{\partial
\alpha}} (J_n\gamma^{\alpha}) - {1\over J_n}
 {\partial\over{\partial\beta}}
(J_n\gamma^{\beta}) -
 {1\over J_n} {\partial\over{\partial
\zeta}} (J_n \gamma^{\zeta})
\end{equation}
where $\gamma^{\alpha}, \gamma^{\beta}$ and 
$\gamma^{\zeta}$ are the fluxes in the $\alpha, 
\beta$ and $\zeta$ directions,
\begin{eqnarray}
\label{no14}
\gamma^{\alpha} & = & - {D_{\alpha\alpha}} {\partial\phi\over{\partial\alpha}} 
 - {D_{\alpha\beta}} {\partial\phi\over{\partial\beta}} 
- {D_{\alpha\zeta}} {\partial\phi
\over{\partial\zeta}}\\
\label{no15}
\gamma^{\beta} & = & - {D_{\beta\alpha}}{\partial\phi\over{\partial\alpha}}
 - {D_{\beta\beta}} {\partial\phi\over{\partial\beta}} 
- {D_{\beta\zeta}} {\partial\phi\over
{\partial\zeta}} \\
\gamma^{\zeta} & = & - {D_{\zeta\alpha}}{\partial\phi\over{\partial\alpha}}
 - {D_{\zeta\beta}} {\partial\phi\over{\partial\beta}} 
 - {D_{\zeta\zeta}} {\partial\phi\over{\partial\zeta}},
\end{eqnarray}
with
\begin{eqnarray}
D_{\alpha\alpha} & = & f^2 D_{ee} + g^2 D_{mm} + 2fg D_{em} \\
D_{\beta\beta} & = & a^2 D_{ee} + b^2 D_{mm} +
      c^2 D_{ss} + 2ab D_{em} + 2ac D_{es} + 2bc D_{ms} \\
D_{\zeta\zeta} & = & p^2 D_{ee} + q^2 D_{mm} + 2pq D_{em} \\
D_{\alpha\beta} & = & D_{\beta\alpha} = af D_{ee} + (bf+ag) D_{em} +
      bg D_{mm} + cf D_{es} + cg D_{ms} \\
D_{\zeta\beta} & = & D_{\beta\zeta} = ap D_{ee} + (bp + aq) D_{em} +
      bq D_{mm} + cp D_{es} + cq D_{ms} \\
D_{\alpha\zeta} & = & D_{\zeta\alpha} = fp D_{ee} + 
gq D_{mm} + (fq + gp) D_{em}.
\end{eqnarray}
The diffusion coefficients are 
\begin{eqnarray}
D_{ee} \equiv {\e}_{\infty}\cdot {\tensor{\bf D}}\cdot {\e}_{\infty};\,\,\,
 D_{em}\equiv {\e}_\infty\cdot {\tensor{\bf D}}\cdot {\m}_\infty;\,\,\,
 D_{mm}\equiv {\m}_\infty\cdot {\tensor{\bf D}}\cdot {\m}_\infty;\label{diffusivity_1}\\
 D_{se} \equiv {\s}_{\infty}\cdot{\tensor{\bf D}}\cdot{\e}_{\infty};\,\,\,
 D_{sm}\equiv {\s}_\infty \cdot {\tensor{\bf D}} \cdot {\m}_\infty;\,\,\,
 D_{ss} \equiv {\s}_{\infty}\cdot{\tensor{\bf D}}\cdot{\s}_{\infty}.\label{diffusivity_2}
\end{eqnarray}
In this set of natural Lagrangian coordinates, 
the anisotropic properties of the metric
tensor are inherited by the diffusive flux in different coordinate directions,
 {\it i.e.},
$\gamma^{\beta} \gg \gamma^{\alpha}$ or $\gamma^{\zeta}.$ This can be illustrated
by considering a chaotic divergence-free flow in which
$\lambda_l^\infty = -\lambda_s^{\infty} = \lambda$ and $\lambda_m^\infty = 0.$
Substituting $g^{ij}=\exp(-2\lambda_l t) {\e} {\e} + \exp(-2\lambda_m t) {\m}{\m}
+ \exp(-2\lambda_s t) {\s}{\s}$ into the tensor diffusivity
and using the orthonormality of ${\e},{\m},$ and ${\s},$  
we find that the diffusion coefficients
satisfy the inequalities
$D_{ee}\approx D_{em}\approx D\exp(-2\lambda t) \ll D_{mm} 
\approx D \le D_{es}\approx
D \sigma_e (D_{ms}\approx D \sigma_m) \ll D_{ss}\approx D\exp(2\lambda t)$ for
$\Lambda_l\approx\Lambda_s^{-1} \gg 1.$ Consequently, 
$D_{\beta\beta} \approx D \exp(2\lambda t)$ is much greater than $D_{\alpha\alpha},
D_{\zeta\zeta}, D_{\alpha\beta}, D_{\zeta\beta},$ and $D_{\alpha\zeta}$ which 
are at most bounded by $D\sigma, \sigma = \sup(\sigma_e, \sigma_s),$ for
$\Lambda_l\approx\Lambda_s^{-1} \gg 1.$

The diffusion in a chaotic flow is one-dimensional.
This remarkable property of equation (\ref{s2e3}) can be illustrated 
by an exact solution for a chaotic flow modeled by a trivial extension
of Arnold's cat map \cite{standardmap}, 
$$
x_{n+1} = x_n + y_n; \,\,\, 
y_{n+1} = x_n + 2 y_n; \,\,\,
z_{n+1} = z_n.
$$
It is easy to check that 
$g^{ij} = \Lambda^{-1}{\e}_\infty{\e}_\infty + {\m}_\infty{\m}_\infty
        + \Lambda {\s}_\infty{\s}_\infty
,\ \lambda=(\ln\Lambda)/{2t}$ a constant, and $f=q=c=1, g=p=a=b=0.$
The diffusion equation (\ref{s2e3}) now takes the simple form
$$
{\partial\phi\over{\partial t}} = - 
D\exp(-2\lambda t) {\partial^2\phi\over{\partial \alpha^2}}
- D {\partial^2\phi\over{\partial \zeta^2}}
- D\exp(2\lambda t) {\partial^2\phi\over{\partial \beta^2}}
$$
This equation can be solved straightforwardly by the method 
of separation of variables.
As an example, for such a flow in an infinitely extended space
an initial distribution of the scalar field 
$\phi(t=0) = c_0(1-\cos k\alpha)(1-\cos k\beta)(1-\cos k\zeta)$
relaxes as 
\begin{eqnarray}
\label{s2e13}
\phi = c_0 & \{ &  1 - \exp[-(1-e^{-2\lambda t})/2\Omega] cos\ k\alpha \}
 \nonumber \\ 
& \{ & 1 - \exp[-(e^{2\lambda t} - 1)/2\Omega] cos\ k\beta \} \nonumber \\
& \{ & 1 - \exp(-D k^2 t) cos k\zeta \}
\end{eqnarray}
with $\Omega = \lambda /{k^2 D}$ the ratio of the characteristic
diffusion time of the passive scalar
($1/{k^2 D}$) and the Lyapunov time of the flow ($1/\lambda$).
One might be concerned that the construction of ${\s}$ lines and finite time
Lyapunov exponents in the example was based on a map but the solution
was given in the continuous time. There are two ways to interpret this result,
neither affects the essential physics. 
One is to regard equation (\ref{s2e13})
as the solution for a time periodic flow field which has the form of the
cat map if sampled at the periods of the flow. 
This is justified since the map and the flow field from which it is
reduced have the same spatial dependence of the 
${\s}$ lines and finite time Lyapunov exponents in Lagrangian coordinates.
In the other approach one simply
interprets equation (\ref{s2e13}) as the diffusive relaxation 
for a map by taking 
$t$ at discrete time intervals. 
We also note that a solution of similar form to 
equation (\ref{s2e13}) was given in
\cite{batchelor} to illustrate the effect of turbulent strains on the 
small scale variation of passive scalars.
 
The solution has distinct characteristic dependence in the 
different coordinate directions.
For $\Omega\gg 1,$ the
function $\phi$ retains its initial $\alpha$ dependence. 
For $t < t_a-1/2\lambda$ with
$t_a\equiv\ln(2\Omega)/{2\lambda}$ the solution is accurately 
approximated by the initial distribution $\phi_0.$ The $\beta$ dependence of
$\phi$ is damped during a short interval 
$1/\lambda$ centered on the time $t=t_a.$
Despite $\phi$ retaining its initial $\alpha$ dependence, 
$(\partial\phi/\partial\alpha)g^{\alpha\alpha}(\partial\phi/\partial\alpha)$
becomes small for $t$ greater than $t_a$ due to the
smallness of the $g^{\alpha\alpha}$ component of the metric tensor. 
The asymptotic form for the gradient of $\phi$ is determined by the slow
varying $\zeta$ dependence, 
$(\nabla\phi)^2 \approx (\nabla\phi_0)^2 \exp(-2t/\tau_d)$
with $\tau_d=1/Dk^2$ the characteristic diffusion time. 
Hence it is no different from that of
an integrable flow.

This can also be shown by examining the rate of the 
production of entropy-like quantity 
$S$  which was defined in equation (\ref{s1e6}).  
In natural Lagrangian coordinates
\begin{eqnarray}
\label{s2e14}
{dS\over{dt}} = \int \Bigl[ {1\over D_{\beta\beta}}(\gamma^{\beta})^2
        & + & \Bigl( D_{\alpha\alpha} - {D_{\alpha\beta}^2\over{D_{\beta\beta}}}\Bigr)
          \Bigl( {\partial\phi\over{\partial\alpha}}\Bigr)^2
        + \Bigl( D_{\zeta\zeta} - {D_{\beta\zeta}^2\over D_{\beta\beta}} \Bigr)
          \Bigl( {\partial\phi\over{\partial\zeta}}
 \Bigr)^2 \nonumber\\
        & + & 2\Bigl( D_{\alpha\zeta} -
          {D_{\alpha\beta} D_{\beta\zeta}\over{D_{\beta\beta}}} \Bigr)
          {\partial\phi\over{\partial\alpha}}{\partial\phi\over{\partial\zeta}}\Big]
        J_n d\alpha d\beta d\zeta.
\end{eqnarray}
The $(\gamma^{\beta})^2$ (the diffusive flux in $\beta$ coordinate) gives the 
main pulse of $S$ production in the time interval $1/\lambda$ centered on the time
$t_a.$ On a longer time scale, this term and the $(\partial\phi/\partial\alpha)^2$
term give an $S$ production that scales as $\exp(-2\lambda t),$ while the
$(\partial\phi/{\partial\zeta})^2$ term makes the dominate contribution 
which scales as $\exp(-2t/\tau_d)$ with $\tau_d=1/Dk^2$ 
the characteristic diffusion time.

In 3D flows the ${\e}_\infty$ and ${\m}_\infty$ vectors are generally mixed in
natural Lagrangian coordinates $\alpha$ and $\zeta.$ Consequently,  diffusion 
in these two coordinate directions are dominated by
 the contribution from ${\m}$ direction and
they have the  characteristic time scale of an integrable flow,
just like the $\zeta$ dependence of $\phi$ in Eq.\ (\ref{s2e13}).
It is the diffusion in the $\beta$
coordinate that distinguishes the transport of a passive scalar in a chaotic
flow from that in an integrable flow.

The ${\s}$ lines give the most important information for constructing 
the natural Lagrangian coordinate system, 
and thus determine the evolution of a passive
scalar. A single ${\s}$ line generically fills a  chaotic region in bounded systems. 
This implies that
the asymptotic ({\it i.e.}, on the time scale which is much longer than the 
typical advection time) evolution of the passive scalar in a generic chaotic flow
is different from that of the simplified solution given earlier, Eq.\ ~(\ref{s2e13}).
That is, the final $\phi$ distribution will not retain any coordinate dependence
in the region where the flow field is chaotic and the smoothing of the gradient
of $\phi$ scales at a rate much faster than $\nabla\phi_0 \exp(-t/\tau_d)$ with
$\tau_d = L^2/D$ the characteristic diffusion time.

It should be noted that the simple model based on cat map is mixing 
in the $x$-$y$ plane and has straight ${\s}$ lines due to hyperbolicity.
Generic flows are only ergodic and can have non-hyperbolic points.
In other words, generic flows can have integrable regions and their ${\s}$ lines
have a complicated geometry. The next two sections study the additional 
features of the properties
of the solution to the advection-diffusion equation which were missing from
the simple model flow based on cat map.

\section{Finite time Lyapunov exponent and barriers for diffusion}
\label{section:exponent}

Unlike the infinite time Lyapunov exponent which is a constant in one
chaotic zone, the finite time Lyapunov exponent for any given time
$\lambda(\xi,t_0)$ can, and generally does, vary significantly 
over space for a generic chaotic flow. The strong spatial dependence of the 
finite time Lyapunov exponent produces a large spread 
in the time during which diffusion is important. 
Such effect can be examined  both  crudely and exactly,
corresponding to a study of the statistical properties and the exact
spatial dependence of the finite time Lyapunov exponent, respectively.

To understand the termination of the enhanced diffusive transport at
a crude level, one can convolute the time $t_a$ with the corresponding
probability distribution function of the finite time Lyapunov exponent.
The probability distribution of the finite time Lyapunov exponents
$\lambda(\xi,t=t_0)\equiv -\ln\Lambda_s(\xi,t=t_0)/2t_0$ 
is approximately Gaussian
with respect to variation in space, so will be the spread in time $t_a.$ 
Since the difference between the distribution of the finite time 
Lyapunov exponents and a Gaussian distribution becomes smaller
as one samples the finite time Lyapunov exponent at a longer time
interval (larger $t_0$), the spread in $t_a$ becomes more Gaussian-like
for systems with longer characteristic diffusion time scale $L^2/D.$
Furthermore, the spread in the time during which the main entropy 
pulse occurs is small if the characteristic diffusion time is long.
This is due to the fact that the standard deviation of the distribution of the 
finite time Lyapunov exponent scales as $1/\sqrt{t_0}.$
Numerical illustration of these properties are given 
in Figs. 2-5 
in section \ref{section:flow}.

A detailed examination of the diffusive transport requires the knowledge
of the exact spatial-temporal dependence of the finite time Lyapunov 
exponent in a given chaotic flow, especially the spatial variation 
of $\lambda(\xi,t)$ along the ${\s}$ lines, since that is the line along
which the rapid diffusive relaxation occurs.
These information are given by equations (\ref{s1e11},\ref{s1e12},\ref{s1e13}).
In \cite{tang} we derived equations (\ref{s1e11},\ref{s1e12},\ref{s1e13})
for two dimensional conservative systems
by applying the constraint that the Riemann-Christoffel curvature 
tensor must vanish in a flat space on which the Lagrangian coordinates
are defined.
A similar calculation in 3D is currently not feasible,
so we instead resort to a numerical resolution.

The key to equations (\ref{s1e11},\ref{s1e12},\ref{s1e13}) is to show
\begin{equation}
\label{s3e1}
\lim_{t\rightarrow\infty} [{\s}_\infty(\xi)\cdot\nabla_0 \lambda(\xi,t) t
+ \nabla_0\cdot{\s}_\infty(\xi)] = 0.
\end{equation}
Once this relationship is established, one can immediately see that
$\lim_{t\rightarrow\infty} {\s}_\infty\cdot\nabla_0\lambda(\xi,t) t$
can not have a time dependence. Let
\begin{equation}
\label{s3e2}
\lim_{t\rightarrow\infty} {\s}_\infty\cdot\nabla_0\lambda(\xi,t) t
= {\s}_\infty\cdot\nabla_0\tilde{\lambda}(\xi)
\end{equation}
with $\tilde{\lambda}$ a time independent smooth function of position.
Equation (\ref{s3e2}) allows a function $f(\xi,t)$ satisfying 
\begin{equation}
\label{s3e3}
{\s}_\infty\cdot\nabla_0 f(\xi,t) = 0
\end{equation}
to be included in the decomposition of $\lambda(\xi,t).$
The function $f(\xi,t)$ is bounded by a $\sqrt{t}$ dependence
in equation (\ref{s1e11}). The obvious
reason is that
$\lim_{t\rightarrow\infty} f(\xi,t)/\sqrt{t}$ has to vanish to satisfy
the definition $\lim_{t\rightarrow\infty} \lambda(\xi,t) = \lambda^\infty.$
The exact choice of $\sqrt{t}$ comes from the fact that the standard 
deviation of the distribution of the finite time Lyapunov exponent 
over space has a
$1/\sqrt{t}$ dependence. Deviation from this $1/\sqrt{t}$ dependence
at finite time is captured by the weak time dependence in $f(\xi,t).$ 
 
We have numerically evaluated 
\begin{equation}
\label{expression}
\Delta(\xi,t) \equiv 
| {\s}\cdot\nabla_0 \lambda_0 t + \nabla_0\cdot{\s} |
\end{equation}
for two different models of three dimensional flows,
section \ref{section:flow}. 
Similar as what we did in \cite{tang}, a finite difference scheme is
avoided by expressing $\Delta(\xi,t)$ in terms of the spatial derivatives
of the metric tensor, appendix \ref{appendix:derivative}.
We find that $\Delta(\xi,t)$ converges exponentially in time to zero.
The convergence rate is approximately equal to that of the ${\s}$ vector,
{\it i.e.}, twice the Lyapunov exponent, 
as can be seen in Fig. 6 in section \ref{section:flow}.
Hence we have numerically validated equation (\ref{s3e1}), which is the 
basis for equations (\ref{s1e11},\ref{s1e12},\ref{s1e13}).

The spatial derivative of the finite time Lyapunov exponent along an ${\s}$
line is proportional to the divergence of the ${\s}_\infty$ vector. 
For straight segments of an ${\s}$ line, the divergence of ${\s}_\infty$
is small, so is the variation in the finite time Lyapunov exponent.
At the sharp bends of an ${\s}$ line, the finite time Lyapunov exponent
makes a large swing in its magnitude in accordance with the large oscillation of 
$\nabla\cdot{\s}_\infty.$  Analytically speaking, the finite time Lyapunov
exponent attains a local minimum along an ${\s}$ line where 
$$
\nabla\cdot{\s}_\infty=0 \,\,\,\,\,\,{\rm and} \,\,\,\,\,\,
{\s}_\infty\cdot\nabla(\nabla\cdot{\s}_\infty) < 0
$$
and reaches a local maximum along an ${\s}$ line where
$$
\nabla\cdot{\s}_\infty=0 \,\,\,\,\,\,{\rm and} \,\,\,\,\,\,
{\s}_\infty\cdot\nabla(\nabla\cdot{\s}_\infty) > 0.
$$
In terms of simple geometry, the finite time Lyapunov exponent has a
maximum where the neighboring ${\s}$ lines are squeezed and 
has a minimum where the neighboring 
${\s}$ lines are bulged outward.  
In the cases that we have studied, the finite time Lyapunov exponent
has a sharp dip at the sharp bends of an ${\s}$ line.
The bending of an ${\s}$ line can be characterized by its local curvature.
In 3D, the curvature of the ${\s}$ lines has an ${\e}$ and an ${\m}$ component.
Figure 8 shows the variation of the finite time Lyapunov exponent
along an ${\s}$ line and the variation of the ${\s}$ line curvature.

The equations (\ref{s1e11},\ref{s1e12},\ref{s1e13}) have a surprisingly broad
range of applications.
It uncovers a direct link between the finite time
Lyapunov exponent and the ${\s}_\infty$ vector field. (Note:
${\s}_\infty$ labels the stable direction, it is 
the tangent vector of the local stable manifold if the later exists. The
${\s}$ line is equivalent to the Lagrangian stable foliation in a general
time dependent flow.)  
By relating geometry (${\s}$ lines) to a dynamical quantity 
(Lyapunov exponent), it provides new insights into the understanding of
chaotic systems in general and hamiltonian systems in particular\cite{tang98}.
The importance of this discovery in transport study is transparent.
It forms the basis for a detailed examination of diffusive transport
in a chaotic flow.
As shown in the last section, the rapid diffusion only occurs along
the ${\s}$ lines. According to equations (\ref{s1e11},\ref{s1e12},\ref{s1e13})
the finite time Lyapunov exponent and hence the characteristic
dimensionless parameter $\Omega$ vary little on a segment of the ${\s}$ lines
which is straight (small $\nabla_0\cdot{\s}_\infty$).
Consequently the spatial gradient
of the passive scalar on a straight ${\s}$ line segment
would be wiped out by a rapid diffusion during a short duration.
The situations are quite different on the two ends of this straight
${\s}$ line segment, which are identified as the sharp bends of the 
${\s}$ line. The finite time Lyapunov exponent has a sharp variation
in its magnitude at these sharp bends of the ${\s}$ lines.
Numerical results have consistently shown a sharp drop in the magnitude
of the finite time Lyapunov exponent, see section \ref{section:flow}.
A peculiarly small ${\lambda}$ leads to a significant reduction
in $\Omega,$ hence a form of local diffusion barrier is created.
A simple analogy is the temperature relaxation in a line of iron rods
bound together by some plastic chips. 
The temperature gradient will be removed in each iron rod very 
quickly but the plastic chips would serve as a practical thermal barrier
on this fast time scale.
Of course, the whole system will reach to thermal equilibrium 
after certain time
if the system is isolated from the surroundings.
The exact time scale for this to happen is given by the thermal 
conductivity of the plastic chips.    
    
The existence of diffusion barriers associated with the sharp bends of
the ${\s}$ lines actually remedies a pathology of 
the natural Lagrangian coordinates in applications. 
The natural Lagrangian coordinates defined in last section 
are intrinsically  local coordinates. Natural Lagrangian coordinates are
 closely related to
the Clebsch coordinates (see appendix \ref{appendix:coordinates}).
It is well known that the  Clebsch coordinates, 
which are also called Euler potentials, are not generally 
single-valued  if one
attempts to extend them over large regions \cite{stern}.
However, this pathology is not  as  important as it first 
appears since the presence of
local diffusion barriers along the ${\s}$ line 
effectively impose boundary conditions
in the natural Lagrangian coordinates, and hence only local coordinates 
are relevant for describing the chaotic transport which has well separated time
scales. 

If not for the second term $f(\xi,t)/\sqrt{t}$ in equation (\ref{s1e11}), 
the finite time Lyapunov exponent would be a smooth function 
in space for arbitrary time.
In fact, the finite time Lyapunov exponent becomes a fractal function
of position across the ${\s}$ lines for large $t,$ 
since $f(\xi,t)$ develops an exponentially growing
spatial gradient in time along directions away 
from the ${\s}_\infty$ direction \cite{tang}. 
       
For any given time $t_0,$ this property is reflected in the correlation
length of the finite time Lyapunov exponent in different directions.
The correlation length along the ${\s}$ line is extremely long since ${\lambda}$
is a smooth function along this direction. 
Across the ${\s}$ line, the irregularity in $f(\xi,t_0)$ overwhelms
the regularity in $\tilde{\lambda}$ and the correlation length for $\lambda$
is greatly reduced. The richest structure and hence the shortest
correlation length, lies along the ${\e}$ lines. 
The fractal nature of function $f(\xi,t)$ brings another degree of complexity
to the diffusive relaxation. That is, the entropy production in a chaotic
flow is a fractal function of space and time.     
In retrospect, the spread in the time during which the main entropy 
production pulse occurs is actually determined by $f(\xi,t),$
since the standard deviation $\sigma(t)$ of the distribution of  
finite time Lyapunov exponents is given by
\begin{equation}
\label{deviation}
\sigma(t) = (\sqrt{\langle f^2 \rangle - {\langle f \rangle}^2} / \lambda^\infty)
t^{-1/2} + O(t^{-1}),
\end{equation}
where $\langle\cdots\rangle$ denotes averaging over space.
 
\section{Flow models and numerical illustration}
\label{section:flow}

To examine the transport problem quantitatively, one has to model the 
chaotic flow. For simplicity, we have used area(volume)-preserving maps to
 model a divergence-free flow. The standard map (SM) \cite{standardmap}
\begin{equation}
\begin{array}{lll}
\displaystyle{x_{n+1}} & \displaystyle{=} & \displaystyle{x_n - {(k/{2\pi})
 }\sin(2\pi y_n)}
\\ \displaystyle{y_{n+1}} &
 \displaystyle{=} & \displaystyle{y_n + x_{n+1},}
\end{array}
\end{equation}
with $k$ a constant, 
is a good choice for modeling a 2D time-periodic divergence-free flow.
We have devised an extended 3D version of the standard map (ESM)
\begin{eqnarray}
x_{n+1} & = & x_n - {(k/{2\pi})} \sin(2\pi y_n) + \Delta \nonumber\\
y_{n+1} & = & y_n - z_n \\
z_{n+1} & = & y_n - x_{n+1} \nonumber
\end{eqnarray}
with $k$ and $\Delta$ constants,
to model a 3D divergence-free flow.
ESM is attractive for studying chaotic advection-diffusion problem since it is
a divergence-free map based on well-studied standard map and 
a point spirals along a KAM surface much the same as the motion of a 
fluid element trapped in a fluid vortex.
 
The ABC flow ${\v}=(v_x, v_y, v_z)$ is another example of 
a three dimensional divergence-free flow \cite{abc_flow,zaslavsky},
\begin{eqnarray}
v_x & = & A \sin z + C \cos y;   \nonumber \\
v_y & = & B \sin x + A \cos z;             \\
v_z & = & C \sin y + B \cos x.   \nonumber
\end{eqnarray}
It satisfies the Beltrami condition $\nabla\times{\v}={\v}$ and allows
chaotic stream lines.
The ABC flow has direct relevance in hydrodynamics since it is a
solution to the Navier-Stokes equation with a forcing term ${\bf F}$
linearly proportional to the velocity field ${\v}$ \cite{zaslavsky_book}.
To increase computational efficiency, we employed a
discretized version of the ABC flow, the so-called ABC map \cite{abc_map},
\begin{eqnarray}
x_{n+1} & = & 
x_n + A \sin z_n + C \cos y_n \,\,\, {\rm mod}(2\pi) \nonumber \\ 
y_{n+1} & = &
y_n + B \sin x_{n+1} + A \cos z_n \,\,\, {\rm mod} (2\pi) \\
z_{n+1} & = &
z_{n} + C \sin y_{n+1} + B \cos x_{n+1}, \,\,\, {\rm mod} (2\pi) \nonumber
\end{eqnarray}
to describe the fluid motion in a three dimensional divergence-free flow.

We find that the eigenvectors of the metric tensor of the Lagrangian 
coordinates converge exponentially in time to their time asymptotic limits
in a chaotic region of the flow. In particular, the ${\s}$ vector converges 
with an exponent of $2\lambda,$ twice the Lyapunov exponent
of the flow. Let $\theta$ and $\varphi$ be  the polar and azimuthal angles
of the ${\s}$ vector in spherical coordinates, 
one finds that $d\theta/dt \propto \exp(-2\lambda t)$ and $d\varphi/dt \propto 
\exp(-2\lambda t),$ Fig. 1.  

In Fig. 2, we show the probability distribution of 
$\lambda(\xi, t=t_0)$ in a single chaotic region.
This distribution is approximately Gaussian,  
but deviations from the Gaussian distribution always occur.
The difference between the finite time Lyapunov exponent distribution and
a Gaussian distribution becomes smaller as 
one samples the finite time Lyapunov exponent
at a longer time interval (longer $t_0$), Fig. 3. 
Here the difference is given by
$residue = \int \|P(x)-P_n(x,1,\sigma)\|dx,$ 
where $x\equiv\lambda(t)/\lambda^\infty,$
$P(x)$ is the distribution function of the finite time Lyapunov exponents and
$P_n(x,1,\sigma)$ is a normal distribution which is centered at $x=1$ and
has the same standard deviation $\sigma$ as 
that of $P(x).$ The standard deviation of 
the distribution of  finite time
Lyapunov exponents decreases if the flow is further 
from being integrable, Fig. 4.
For larger $t_0$ (compared with the Lyapunov time) 
the standard deviation of the 
distribution of  finite time Lyapunov exponents 
scales as $1/\sqrt{t_0},$ Fig. 5.

We evaluate $\Delta(\xi,t)$ defined by equation (\ref{expression}) for 
both the extended standard map and the ABC map, figure 6.
It is easy to see that $\Delta(\xi,t)$ converges exponentially in time,
a result that is essential to establish equations 
(\ref{s1e11},\ref{s1e12},\ref{s1e13}), section \ref{section:exponent}.
The strong spatial variation and anisotropy of $f(\xi,t)$ in equation (\ref{s1e11})
are illustrated in figure 7.
One can see that the correlation length is remarkably long along the ${\s}$ lines,
while it is extremely short in directions away from this orientation.
The correlation length of the finite time Lyapunov exponent
along the ${\e}_\infty$ represents the characteristic correlation 
length in a chaotic flow.
The variation of the finite time Lyapunov exponent along an ${\s}$ line
is examined again in figure 8. The geometry of the ${\s}$ line is represented
by the ${\e}$ and ${\m}$ components of the ${\s}$ line curvature.
It is easy to see that there is a sharp dip in the magnitude of the finite 
time Lyapunov exponent wherever the ${\s}$ line makes a sharp bend. 
Peculiarly small finite time Lyapunov exponent leads to small local $\Omega$ 
number and gives rise to effective diffusion barriers.

\section{Transport in an integrable region of the flow}
\label{section:integrable}

In an integrable region of a divergence-free flow, 
neighboring fluid points separate
(or converge) at most algebraically. Consequently, 
the largest eigenvalue $\Lambda_l$
(or the smallest eigenvalue $\Lambda_s$) of the metric tensor of the Lagrangian
coordinates increases (or decreases) at most algebraically. 
The eigenvectors of the metric tensor still have well-defined time
asymptotic limits. Hence the natural Lagrangian coordinates 
introduced in last section
are well-defined in the integrable regions of a flow.

This can be illustrated by considering a divergence-free flow
in a bounded integrable region.
If there is no null point in the region of interest, a
globally divergence-free field admits a Hamiltonian structure \cite{boozer_enc}
to which the machineries in hamiltonian mechanics can be applied. 
Hence the integrable region of such a divergence-free flow consists of bounded
constant ``action'' surfaces \cite{arnold}, the KAM surfaces.
Parameterizing the integrable surfaces using ``action'' implies
the existence of an ``action'' function $\Psi({\x})$
such that ${\v}\cdot\nabla{\Psi}=0$ 
with $\|\nabla\Psi\|\not= 0.$ 
Since the flow is also divergence-free ($\nabla\cdot{\v} = 0$), 
one can treat it as a
one degree of freedom, time dependent Hamiltonian system and write the 
flow field in the canonical representation, in analogy to the canonical 
representation of the magnetic field\cite{boozer_coordinate}. That is
\begin{equation}
\label{s4e1}
{\v} = \nabla\Psi \times \nabla\Theta + \nabla\Phi \times \nabla\chi(\Psi)
\end{equation}
with the hamiltonian $\chi$  a function of the action-like quantity $\Psi$ alone.
The motion of the fluid element in the $\Phi$ coordinate is determined by the 
Jacobian $J$ of the $\Psi$-$\Phi$-$\Theta$ coordinates,
\begin{equation}
\label{s4e2}
d\Phi/dt = {\v}\cdot\nabla\Phi = (\nabla\Psi\times\nabla\Theta)\cdot\nabla\Phi 
= \nu(\Psi,\Phi,\Theta) = 1/J,
\end{equation}
The Jacobian $J$ is in general a function of all three coordinates.
The angle-like variables $\Theta$ and time-like variable $\Phi$ 
are periodic and we set the period to be $2\pi.$
The topology of the flow trajectory on a KAM surface is simple 
in canonical coordinates, and it is given 
by $\Theta = \Theta_0 + \iota(\Psi) \Phi$ with $\iota=d\chi(\Psi)/d\Psi$ 
the winding number of the flow trajectory.
By good KAM surfaces, we mean the surfaces that have irrational
winding numbers. Surfaces with a rational winding number consist of closed lines. 
But a similar line can not close on itself on a good KAM surface.

Straightforward substitution shows that the transformation of
 $\Phi = \varphi + \varrho, \ \Theta=\vartheta + \iota \varrho,$
 gives the 
same flow field ${\v}$ in canonical form, Eq.\ (\ref{s4e1}).
 Except for the arbitrary function $\varrho,$ the $\Psi$-$\Phi
(\phi)$-$\Theta(\theta)$
coordinates are uniquely defined.
This limited arbitrariness, in return, allows one to make a transformation of 
 $\Phi\rightarrow\varphi$ and $\Theta\rightarrow\vartheta,$
such that the motion of the fluid element on a good KAM surface is prescribed by
\begin{equation}
\label{s4e3}
\Psi = \Psi_0; \hspace{0.5cm}
\varphi = \varphi_0 + \nu_0(\Psi) t; \hspace{0.5cm}
\vartheta = \vartheta_0 + \iota(\Psi) \nu_0(\Psi) t.
\end{equation}

To prove the existence of equation (\ref{s4e3}), we need to show 
that there exists  a function
$\varrho(\Psi,\Phi,\Theta)$ such that the Jacobian of the new coordinates
$\Psi$-$\varphi$-$\vartheta$ is a function of $\Psi$ alone, {\it i.e.},
$(\nabla\Psi\times\nabla\vartheta)\cdot\nabla\varphi = \nu_0(\Psi).$
Expressing the Jacobian of the $\Psi$-$\Phi$-$\Theta$ coordinates 
in terms of the new 
Jacobian (function of $\Psi$ only) of the transformed coordinates and
the transformation function $\varrho,$ one has
\begin{equation}
\label{s4e4}
{\partial\varrho\over{\partial\varphi}} + 
\iota {\partial\varrho\over{\partial\vartheta}}
= {{\nu(\Psi, \varphi,\vartheta) - \nu_0}\over\nu_0}.
\end{equation}
The double periodicity in $\varphi$ and $\vartheta$ implies that a scalar function
$\nu$ can be written as
\begin{equation}
\label{s4e5}
\nu = \sum_{nm} \nu_{nm} \exp[i(n\varphi - m\vartheta)],
\end{equation}
and the transformation function $\varrho$ can be written as
\begin{equation}
\label{s4e6}
\varrho = \sum_{nm} \varrho_{nm} \exp[i(n\varphi - m\theta)].
\end{equation}
It is easy to show that the Fourier components of the 
transformation function $\varrho$ are
$\varrho_{nm} = i\nu_{nm}/(n - \iota m)\nu_0.$
The desired Jacobian of the new coordinates $\Psi$-$\varphi$-$\vartheta,$ 
$1/\nu_0(\Psi),$ is the $m=0,\ n=0$ Fourier component of
 $\nu(\Psi, \varphi,\vartheta),$ {\it i.e.}, $\nu_0(\Psi) = \nu_{00}.$
This proves the existence of $\Psi,\ \varphi,\ \vartheta$ coordinates in which the
motion of fluid element on a good KAM surface satisfies equation (\ref{s4e3}). 

If the system is perturbed away from complete integrability, there exist
remnant KAM surfaces which are parameterized on a discontinuous set of action.
The trajectory on a KAM surface still follows equation (\ref{s4e3}) but the 
action coordinate is generally on a Cantor set. P{\"o}schel showed that on
this Cantor set, the KAM surfaces form a differentiable family in the sense of
Whitney so one can speak of an integrable system on a Cantor set \cite{poschel}.
The construction of the metric tensor needs the $\Psi$ derivative of the 
Jacobian $\nu_0$ and the rotational transform $\iota.$ By following P{\"o}schel,
the $\Psi_0$ derivative of $\nu_0$ and $\iota$ can be properly defined (in the 
sense of Whitney) on the remnant KAM surfaces. 
Except for this subtlety, the results presented in the next two paragraphs
on the properties of the metric tensor applies to the remnant KAM surfaces
in a perturbed system.

In $\Psi$-$\varphi$-$\vartheta$ coordinates, the Jacobi matrix of the Lagrangian 
coordinates $\Psi_0$-$\varphi_0$-$\vartheta_0$ is simple,
\[
\tensor{\bf J} = \left(
\begin{array}{ccc}
1 & 0 & 0 \\
{\cal A} t & 1 & 0 \\ 
{\cal B} t & 0 & 1 
\end{array}
\right)
\]
with $ {\cal A} = \partial\nu_0/\partial\Psi $ and
 $ {\cal B} = \partial(\iota\nu_0)/\partial\Psi .$
Without losing  generality, we write the metric tensor of the 
$\Psi$-$\varphi$-$\vartheta$ coordinates as
\[
\tensor{\bf g}_0 = \left(
\begin{array}{ccc}
{\cal C} & {\cal D} & {\cal E} \\
{\cal D} & {\cal F} & {\cal G} \\
{\cal E} & {\cal G} & {\cal H}
\end{array}
\right)
\]
where ${\cal C, D, E, F, G}$ and ${\cal H}$ are the covariant components $g_{ij}$
of the metric tensor $\tensor{\bf g}_0.$
The determinant of $\tensor{\bf g}_0$ in covariant representation is 
\begin{equation}
\label{s4e7}
\|\tensor{\bf g}_0\| \equiv J_0^2 \equiv {\cal C}{\cal F}{\cal H} + 2{\cal D}{\cal E}{\cal G}
        - {\cal F}{\cal E}^2 - {\cal C}{\cal G}^2 - {\cal H}{\cal D}^2.
\end{equation} 
The metric tensor of the Lagrangian coordinates $\Psi_0$-$\varphi_0$-$\vartheta_0$
is given by
\begin{equation}
\label{metric}
 \tensor{\bf g} = \tensor{\bf J}^{\rm T}\cdot\tensor{\bf g}_0\cdot\tensor{\bf J}
\end{equation}
with $\tensor{\bf J}^{\rm T}$  the transpose of $\tensor{\bf J}.$

For large $t,$ the three eigenvalues of the metric tensor
 ($\tensor{\bf g}$) of the Lagrangian
coordinates $\Psi_0$-$\varphi_0$-$\vartheta_0$
are given by
\begin{eqnarray}
\label{s4e8}
\Lambda_l & = & ({\cal F}{\cal A}^2+{\cal H}{\cal B}^2 + 2{\cal G}{\cal A}{\cal B}) t^2
        + 2 ({\cal D}{\cal A}+{\cal E}{\cal B}) t + {\cal O}(1)  \nonumber \\
\Lambda_m & = & {{ {\cal H}{\cal F}{\cal A}^2 + {\cal F}{\cal H}{\cal B}^2
        - {\cal G}^2{\cal B}^2 - {\cal G}^2{\cal A}^2} \over {
        {\cal F}{\cal A}^2+{\cal H}{\cal B}^2 + 2{\cal G}{\cal A}{\cal B} }}
        + {\cal O}(t^{-1}) \\
\Lambda_s & = & {{J_0^2}\over{{\cal H}{\cal F}{\cal A}^2 +
         {\cal F}{\cal H}{\cal B}^2
        - {\cal G}^2{\cal B}^2 - {\cal G}^2{\cal A}^2}} {1\over t^2} +
         {\cal O}(t^{-3}). \nonumber
\end{eqnarray}
The three eigenvectors converge linearly in time to their asymptotic limits,
${\e}_\infty = (1, 0, 0),\ {\m}_\infty \propto (0, \iota'\nu_0 + \iota\nu_0', -\nu_0'),\
 {\s}_\infty \propto (0, \nu_0', \iota'\nu_0 + \iota\nu_0'),$ 
see appendix \ref{appendix:convergence}. 
Here the prime denotes a derivative with respect to $\Psi.$
The ${\e}$ line is perpendicular to the good KAM surfaces while the ${\s}$ line and 
${\m}$ line always lie on  a good KAM surface.
One ${\s}$ line or one ${\m}$ line generically fills the whole surface, 
as they do in a single chaotic region.

If $\nu_0$ in equation (\ref{s4e3}) is a constant, 
the flow is effectively two dimensional.
The corresponding ${\e}$ and ${\s}$ lines
coincide with the action-like variable and the angle-like variable axes. 
This zero shear case is equivalent to the two dimensional twist map
(standard map at k=0). 

The diffusion coefficients defined in Eqs.\, (\ref{diffusivity_1},\ref{diffusivity_2})
 can be found exactly,
\begin{eqnarray}
D_{ee} & = & c_0 D; \\
D_{em} & = & c_1 D; \\
D_{es} & = & - c_0 D \varpi t + c_2 D; \\
D_{mm} & = & c_3 D; \\
D_{ms} & = & - c_1 D \varpi t + c_4 D; \\
D_{ss} & = & c_0 D \varpi^2 t^2 - c_2 D \varpi t + c_5 D,
\end{eqnarray}
where $c_i,\ i=0,5$ are time independent functions of ${\cal A,B,C,D,E,F,G,H,}$
and their explicit forms are given in appendix \ref{appendix:diffusivity}.
The shearing rate of the flow is $\varpi \equiv \sqrt{{\cal A}^2 + {\cal B}^2}.$ 
The shearing time $1/\varpi$ is the characteristic time of 
a nontrivial integrable flow.

For $t$ large compared with the shearing time $1/\varpi,$
the tensor diffusivity is highly
anisotropic, $D_{ee} \approx D_{mm} \approx D_{em} \approx D \ll \| D_{es} \| \approx
\| D_{ms} \| \approx D\varpi t \ll D_{ss} \approx D\varpi^2 t^2.$
Hence there are  fast diffusion and slow diffusion directions in an integrable 
flow with shear ($\nu$ or $\iota$ is a function of $\Psi$ instead of a constant).
The effective diffusivity in ${\s}_\infty$ direction increases quadratically in time,
so  there is a fast diffusion  along the ${\s}_\infty$ lines,
which lie on the KAM surface.
The ${\e}_\infty$ vector is perpendicular to the KAM surfaces and the effective
diffusivity in ${\e}_\infty$ direction is the classical diffusivity $D.$ 
Hence the diffusion across the KAM surfaces is slow. 
The natural Lagrangian coordinates defined by equation (\ref{s2e1}),
separate these different diffusion time scales and give the 
general properties of the passive scalar transport in a generic integrable flow.
Since $\| D_{es} \| \approx D\varpi t$ and the diffusive flux across KAM surfaces
has the form of equation (\ref{no14}), there is a period of 
enhanced diffusive flux across the KAM surfaces during the time in which the fast 
diffusion is accomplished on the KAM surfaces.  

These transport properties can also be demonstrated 
by solving the diffusion equation (\ref{s1e3})
with the rough approximations that
${\cal C} ={\cal F} = {\cal H} =1$ and ${\cal D}={\cal E}={\cal G}=0.$
The metric tensor in its contravariant component ($g^{ij}$), 
now takes the form
\[
\tensor{\bf g} = \left(
\begin{array}{ccc}
1 & -{\cal A} t & - {\cal B} t \\
-{\cal A} t & 1 + {\cal A}^2 t^2 & {\cal A}{\cal B} t^2 \\
-{\cal B} t & {\cal A}{\cal B} t^2 & 1+{\cal B}^2 t^2
\end{array}
\right)
\]
For constant ${\cal A}$ and ${\cal B},$ 
the diffusion equation can be written as
\begin{eqnarray}
{\partial\phi\over{\partial t}} = & & D {\partial^2\phi\over{\partial\Psi_0^2}}
        - 2D{\cal A}t{\partial^2\phi\over{\partial\Psi_0\partial\varphi_0}}
        - 2D{\cal B}t{\partial^2\phi\over{\partial\Psi_0\partial\vartheta_0}}
	+ 2D{\cal A}{\cal B} t^2 {\partial^2\phi\over{\partial\varphi_0\partial\vartheta_0}} \nonumber \\
        & &+ D(1 + {\cal A}^2 t^2){\partial^2\phi\over{\partial\varphi_0^2}}
        + D(1 + {\cal B}^2 t^2){\partial^2\phi\over{\partial\vartheta_0^2}}. \label{s4e9}
\end{eqnarray}
The general solution to this equation is
\begin{equation}
\label{s4e10}
\phi(\Psi_0,\varphi_0,\vartheta_0,t) = (2\pi)^{-3/2}
        {\int\!\!\int\!\!\int}_{-\infty}^{\infty} 
        \tilde{\phi}(k_{\Psi},k_{\varphi},k_{\vartheta},t)
        e^{i(k_{\Psi} \Psi_0 + k_{\varphi} \varphi_0 + k_{\vartheta} \vartheta_0)}
        dk_\Psi dk_\varphi dk_\vartheta
\end{equation}
where
\begin{eqnarray}
\label{s4e11}
\tilde{\phi}(k_{\Psi},k_{\varphi},k_{\vartheta},t)
        & = & \tilde{\phi}_0(k_{\Psi},k_{\varphi},k_{\vartheta})
        e^{-D k_{\Psi}^2 t + {\cal A}t^2 D k_{\Psi} k_{\varphi} + 
         {\cal B}t^2 D k_{\Psi} k_{\vartheta} - 
	(2{\cal A}{\cal B} t^3/3) D k_{\varphi} k_{\vartheta} - 
        (t + {\cal A}^2 t^3/3) D k_{\varphi}^2  -
        (t + {\cal B}^2 t^3/3) D k_{\vartheta}^2} \\
\label{s4e12}
& = &  \tilde{\phi}_0(k_\Psi,k_\varphi,k_\vartheta)
        e^{-Dt [k_\Psi^2 + k_\varphi^2 + k_\vartheta^2
	- {\cal A}t k_\Psi k_\varphi - {\cal B}t k_\Psi k_\vartheta
	+ ({\cal A}t k_\varphi + {\cal B}t k_\vartheta)^2/3]}
\end{eqnarray}
with $\tilde{\phi}_0(k_\Psi,k_\varphi,k_\vartheta)$ given by the initial condition
 $\phi_0\equiv\phi(\Psi_0,\varphi_0,\vartheta_0,t=0)$
\begin{eqnarray}
\label{s4e13}
\tilde{\phi}_0(k_\Psi,k_\varphi,k_\vartheta) & \equiv & 
        \tilde{\phi}(k_\Psi,k_\varphi,k_\vartheta, t=0) \nonumber \\
& = &  (2\pi)^{-3/2} {\int\!\!\int\!\!\int}_{-\infty}^{\infty}
        \phi_0(\Psi_0,\varphi_0,\vartheta_0)
         e^{i(k_\Psi \Psi_0 + k_\varphi \varphi_0 + k_\vartheta \vartheta_0)}
        d\Psi_0 d\varphi_0 d\vartheta_0.
\end{eqnarray}

Let $\tau_{\parallel}$ and $\tau_{\perp}$ be the characteristic diffusion times of the 
initial passive scalar field in and across the KAM surfaces. 
The characteristic dimensionless quantity is the ratio 
between the characteristic diffusion
time and the shearing time of the flow, $\Omega\equiv\varpi\tau_{\parallel}.$
For $\Omega \gg 1,$ the scalar field is advected by the flow until time
$t_a \equiv \Omega^{1/3} /\varpi,$ 
which is much shorter than the characteristic diffusion
time of the initial scalar field.
In the KAM surface, the spatial dependence ($\varphi_0$ and $\vartheta_0$)
of the passive scalar field $\phi$
is damped after another $t_a,$ {\it i.e.}, $\partial\phi/\partial\varphi_0 
\approx \partial\phi/\partial\vartheta_0 \approx 0$ for $t> 2t_a.$
The asymptotic form for the passive scalar field $\phi$ is determined by the 
slow varying $\Psi_0$ dependence,{\it i.e.}, for $t > 2t_a,$
\begin{equation}
\label{s4e14}
\phi(\Psi_0,t) = (2\pi)^{-1/2} {\int}_{-\infty}^{\infty}
        \tilde{\phi}_0(k_\Psi,0,0) e^{i k_\Psi \Psi_0 - D k_\Psi^2 t}dk_\Psi,
\end{equation}
with
\begin{equation}
\label{s4e15}
\tilde{\phi}_0(k_\Psi,0,0) = (2\pi)^{-3/2} {\int\!\!\int\!\!\int}_{-\infty}^{\infty}
        \phi_0(\Psi_0,\varphi_0,\vartheta_0) e^{i k_\Psi \Psi_0} 
        d\Psi_0d\varphi_0 d \vartheta_0,
\end{equation}
and $\phi_0(\Psi_0,\varphi_0,\vartheta_0)$ the initial field.
Hence the smoothing of the gradient of $\phi$ across the KAM surfaces
has a long tail and is accurately described by the characteristic diffusion
time $\tau_{\perp},\ \partial\phi/\partial\Psi_0 \propto \exp(-t/\tau_{\perp}).$

In summary, the fast diffusion which is the result of shearing between
different KAM surfaces and the constraint of the flow being divergence-free, occurs 
only within the KAM surfaces. Diffusion across the KAM surfaces 
is approximated by the characteristic diffusion time and  is very slow.
In the case of the temperature of electrons confined on good magnetic surfaces
 in fusion devices,
the electron temperature quickly relaxes to thermal equilibrium on the good magnetic 
surfaces, while the heat transfer across magnetic surfaces 
is much slower and described 
by a cross-field thermal diffusion time.
On the contrary,  electron temperature variations are rapidly damped in the region of
stochastic field lines, since an ${\s}$ line in which 
there is a rapid diffusion, generically
fills the whole region explored by the stochastic field line.

\section{Summary}
The advection and diffusion of a passive scalar have been investigated
in both chaotic and integrable flows.
The characteristic time scale of a chaotic flow is the Lyapunov time
which measures the exponential convergence of neighboring fluid elements.
The characteristic dimensionless quantity for the chaotic transport problem is
the ratio between the characteristic diffusion time of 
the scalar field and the Lyapunov
time of the flow. This number is in general very large.
The scalar field is purely advected by the flow until the time 
$t_a-1/2\lambda$ with $t_a\equiv\ln2\Omega/{2\lambda}.$
There is a rapid diffusion during a relatively short interval ($1/\lambda$)
centered on time $t_a.$
This rapid diffusion occurs only along the field line of the ${\s}_\infty,$
which defines the stable direction for the streamlines. 
The fast diffusion can be
confined to one coordinate in natural Lagrangian coordinates.
The rapid diffusion removes the gradient of the scalar field in the entire
chaotic region.

The finite time Lyapunov exponent varies smoothly along an ${\s}$ line
and has sharp dips where the ${\s}$ line makes a sharp bend.
A large reduction in $\lambda$ leads to a peculiarly small $\Omega$ number.
Hence the sharp bends of the ${\s}$ line define a class of barriers
for diffusion. This new class of diffusion barriers 
are associated with the non-hyperbolicity of the system, which
is thought to be generic for chaotic systems \cite{tang}.

The characteristic time scale of an integrable flow with shear is the  
time scale on
which neighboring fluid points separate algebraically due to the shear.
The characteristic dimensionless quantity for the transport of a 
passive scalar in such 
flow is the ratio between the characteristic diffusion time of 
the scalar field 
and the shearing time of the flow, $\Omega \equiv \tau_d\varpi.$
If the shearing time of the flow is much faster than the 
characteristic diffusion
time, the scalar field is advected by the flow until time 
$t_a \equiv  \Omega^{1/3}/\varpi.$
The fast diffusion, which is confined within the KAM surfaces, removes the gradient
of the scalar field in the KAM surfaces after time interval $t_a.$
During the  period ($t_a<t<2t_a$), there is an enhanced diffusive flux
(compared with the one predicted by the characteristic diffusion time) 
across the KAM surfaces, but it is too small to remove the $\Psi_0$ dependence.
For $t>2t_a,$ the scalar field has only $\Psi$ dependence, and its decay is 
accurately described by the characteristic diffusion time.
Hence, across the 
KAM surfaces, the diffusion is distinctly slow and  a large gradient of
the scalar field can be maintained.

\acknowledgments
We would like to thank U. S. Department of Energy for support under grant
DE-FG02-97ER54441. Part of the research was done while one of the authors
(Tang) was supported by a NSF University-Industry Postdoctoral Fellowship
in Mathematical Sciences through SUNY Stony Brook.

\appendix

\section{Chaotic flow and the ergodic theorem of dynamical systems}
\label{appendix:ergodic}

If a flow field is smooth, 
the equation of motion for the fluid element, Eq.\ (\ref{s1e2}),
can be treated as a differentiable dynamical system to which the 
ergodic theorem of dynamical systems \cite{ruelle85} can be applied.

For simplicity, we consider a steady flow, $d{\x}/dt = {\v}({\x}),\,
{\x}\in\Re^3,$ or a time-periodic flow which can be reduced to a map,
${\x}_{n+1} = {\V}({\x}_n),\ {\x}_n\in\Re^3.$
We also assume that the flow is time reversible.
The distance between neighboring points
at time $t$ is related to their initial separation
by $dl^2 = g_{ij}d\xi^id\xi^j,$ with $g_{ij}$ 
the metric tensor of the Lagrangian
coordinates.  
The rate of the exponential divergence or convergence of  neighboring
trajectories is measured by the Lyapunov exponent,
 $\lambda=\lim_{t\rightarrow\infty} (1/2t)
\ln (dl^2/dl_0^2).$
In vector form,
\begin{equation}
\lambda(\xi,{\u}) = \lim_{t\rightarrow\infty}
 \ln({{\u}\cdot\tensor{\bf g}\cdot{\u}}) /{2t}.
\end{equation}
Here ${\u}$ specifies the direction along 
which the initial fluid points separate,
{\it i.e.}, ${\bf{\bar \delta_0}} = \delta_0 {\u}.$

In a single chaotic region (the region in which an ergodic measure is preserved 
by the time evolution of the fluid equation), the multiplicative ergodic theorem 
asserts that there exist three characteristic directions in which three 
Lyapunov exponents reside, {\it i.e.},
\begin{equation}
\lambda_i = \lambda(\xi,{\e}^i) = \lim_{t\rightarrow\infty} \ln({{\e}^i\cdot
\tensor{\bf g}\cdot{\e}^i}) / {2t}, \hspace{0.5cm} i=1,2,3.
\end{equation}
The Lyapunov exponents are independent of position $\xi$ in a single chaotic region.
If there is no degeneracy in Lyapunov exponents, $\lambda_1>\lambda_2>\lambda_3,$
which is trivially true for a chaotic divergence-free flow,
the three-dimensional basis ${\e}^i,\ i=1,2,3,$  which are functions of Lagrangian 
coordinates alone, are distinct and span $\Re^3.$
For a rigorous mathematical proof, see \cite{oseledec} and \cite{ruelle}.
For a general discussion, see \cite{standardmap}. 
Generically, ${\e}^i,\ i=1,2,3,$ are not orthogonal to each other.  
The eigenvectors of the metric tensor of the Lagrangian coordinates  are
 orthogonal to each other, and their time asymptotic limits 
are uniquely related to the characteristic directions ${\e}^i$
by
\begin{equation}
{\e}_\infty \propto {\e}^2\times{\e}^3;\hspace{0.5cm} 
{\m}_\infty \propto {\e}^2 - ({\e}^2\cdot{\e}^3){\e}^3;\hspace{0.5cm}
{\s}_\infty = {\e}^3.
\end{equation}
The finite time eigenvectors converge exponentially to their time asymptotic
limit, Fig. 1 in section \ref{section:flow}.
 
In an integrable region of the flow, the Lyapunov exponents vanish.
But for a nontrivial flow (flow with shear), there exist
non-degenerate characteristic directions which are associated with the center
unstable, center, and center stable manifolds \cite{ruelle_book}. 
Hence the eigenvectors of the metric tensor still have well-defined time
asymptotic limits, but with an algebraic convergence rate, as we showed
in section \ref{section:integrable}.

\section{Advection and diffusion in a flow with 
${\protect \lambda_{\protect m}^\infty\neq 0}$}
\label{appendix:3D}

The approach presented in this paper can be applied to flows with an
arbitrary combination of positive and negative Lyapunov exponents.
The trajectory of a flow point, which is the solution to equation
$d{\x}/dt = {\v}({\x},t),$ is characterized by at most three Lyapunov 
exponents. For a general time-dependent divergence-free flow, there are always 
one positive ($\lambda_l>0$) and
one negative ($\lambda_s<0$). The middle one $\lambda_m$ might be non-zero.
If $\lambda_m>0,$ the effective diffusivity in ${\m}_\infty$ direction
$$
D_{mm}\equiv {\m}_\infty\cdot\tensor{D}\cdot{\m}_\infty 
\approx D/\exp(2\lambda_m t)
$$
decreases exponentially in time, just like that in the 
${\e}_\infty$ direction. Consequently, diffusion occurs only along
the field line of the ${\s}_\infty$ vector. 

Even if $\lambda_m<0,$ the rapid diffusion in a chaotic flow occurs
only along the ${\s}$ line, as long as $\lambda_m$ does not have a 
value very close to that of $\lambda_s.$ 
This can be seen by comparing the effective diffusivities in ${\m}_\infty$
and ${\s}_\infty$ directions at time $t_a\equiv \ln(2\Omega)/2|\lambda_s|$
with $\Omega\equiv |\lambda_s| L^2/D,$
$$
{D_{mm}\over {D_{ss}}} \approx
\exp[2(|\lambda_m| - |\lambda_s|) t_a] 
= (2\Omega)^{-1+ |\lambda_m|/|\lambda_s|}.
$$
For $\Omega\gg 1$ which is the case for most practical problems,
$D_{mm}/D_{ss} \ll 1$ if $\lambda_m\neq\lambda_s.$
That is, the diffusion occurs only along the ${\s}$ line.
If $\lambda_m=\lambda_s,$ diffusion occurs in the $({\m}_\infty,{\s}_\infty)$
surfaces and diffusion barriers appear where both $\lambda_m$ and $\lambda_s$
have peculiarly small values.

\section{Construction of natural Lagrangian coordinates}
\label{appendix:coordinates}
If ${\s}_\infty$ is an arbitrary vector field, one can find a function $g(\xi)$
such that $e^{-g(\xi)} {\s}_\infty$ is divergence free, 
for $\nabla\cdot(e^{-g(\xi)} {\s}_\infty)
= (- {\s}_\infty \cdot \nabla g(\xi) + \nabla\cdot{\s}_\infty) e^{-g(\xi)}$
can be made to vanish by solving for $g(\xi)$ such that
 ${\s}_\infty\cdot\nabla g=\nabla\cdot{\s}_\infty.$
Divergence-free fields can be represented in Euler potentials \cite{euler}
 $\alpha$ and $\zeta,$
{\it i.e.}, $e^{-g(\xi)} {\s}_\infty = \nabla\zeta\times\nabla\alpha$
Hence
an arbitrary field ${\s}_\infty$ can be written in the Clebsch representation,
\begin{equation}
\label{a1}
{\s}_\infty = e^{g({\xi})} \nabla\zeta\times\nabla\alpha
\end{equation}
where the Euler potentials $\alpha({\xi})$ and $\zeta({\xi})$ are
 locally defined functions such that
${\s}_{\infty} \cdot \nabla\alpha = 0$ and ${\s}_{\infty} \cdot \nabla\zeta = 0.$

Using the dual relations \cite{boozer_enc}, we can relate the third coordinate $\beta$
to the ${\s}_\infty$ field, {\it i.e.},
\begin{equation}
\label{a2}
{\s}_\infty = { e^{g({\xi})} \over J}{\partial {\xi}\over{\partial \beta}}
\end{equation}
with $J = 1/{(\nabla\alpha\times\nabla\beta)\cdot\nabla\zeta}$ the Jacobian of
the $\alpha$-$\beta$-$\zeta$ coordinates. The choice of the Jacobian is free.
 One can let $J=1$ or $J= e^{g({\xi})}.$

Since ${\e}_\infty = {\m}_\infty\times{\s}_\infty$ and ${\m}_\infty =
{\s}_\infty\times{\e}_\infty,$ one can show
\begin{eqnarray}
\label{a4}
{\e}_\infty & = & - ({\m}_\infty\cdot\nabla\zeta)  e^{g({\xi})} \nabla\alpha
+ ({\m}_\infty\cdot\nabla\alpha)  e^{g({\xi})} \nabla\zeta\\
\label{a5}
{\m}_\infty & = & ({\e}_\infty\cdot\nabla\zeta) e^{g({\xi})} \nabla\alpha
- ({\e}_\infty\cdot\nabla\alpha)  e^{g({\xi})} \nabla\zeta.
\end{eqnarray}
The ${\s}_\infty$ vector can be written in the general covariant form,
\begin{equation}
\label{a6}
{\s}_\infty = a_1\nabla\alpha + a_2\nabla\beta + a_3\nabla\zeta,
\end{equation}
where only $a_2$ is constrained by $a_2 = J/ e^{g({\xi})}.$
Equations (\ref{a4})-(\ref{a6}) have the required form to yield equation (\ref{s2e1})
 of the paper. It is interesting to note that one choice of $g(\xi)$ is 
$\tilde{\lambda}$ in equation (\ref{s1e11}).

\section{The eigenvectors of the metric tensor}
\label{appendix:metric}
For a general three dimensional flow, one has 
\begin{equation}
\label{c_1}
g_{ij} = \Lambda_l e_i e_j + \Lambda_m m_i m_j + \Lambda_s s_i s_j
\end{equation}
and
\begin{equation}
\label{c_2}
g^{ij} = E^i E^j / \Lambda_l + M^i M^j / \Lambda_m + S^i S^j / \Lambda_s
\end{equation}
with the eigenvalues $\Lambda_l \ge \Lambda_m \ge \Lambda_s > 0.$
Here $e_i,\,m_i,\,s_i$ are the covariant components of the vectors
${\e},\,{\m},\,{\s},$ while $E^i,\,M^i,\,S^i$ are the contravariant
components of the vectors ${\E},\,{\M},\,{\S}.$
They satisfy the relations:
$e_i e_j + m_i m_j + s_i s_j = \delta_{ij};\,
E^i E^j + M^i M^j + S^i S^j = \delta^{ij};\,
\sum e_i E^i = \sum m_i M^i = \sum s_i S^i =1;\,
\sum e_i M^i = \sum e_i S^i = \sum m_i E^i = \sum m_i S^i = \sum s_i E^i
= \sum s_i M^i = 0.$
In vector form, that is:
${\e}\cdot{\E}={\m}\cdot{\M}={\s}\cdot{\S}=1$ and
${\e}\cdot{\M}={\e}\cdot{\S}={\m}\cdot{\E}={\m}\cdot{\S}={\s}\cdot{\E}
={\s}\cdot{\M} = 0.$

To find the dot product of two vectors both of which are in the same form
(covariant or contravariant), one has to specify the metric tensor.
In real space, the metric tensor of the Lagrangian coordinates is
given in equations (\ref{c_1},\ref{c_2}), hence one has 
${\e}\cdot{\e} = \sum e_i g^{ij} e_j = 1/\Lambda_l,\,
{\m}\cdot{\m} = \sum m_i g^{ij} m_j = 1/\Lambda_m,\,
{\s}\cdot{\s} = \sum s_i g^{ij} s_j = 1/\Lambda_s,\,
{\E}\cdot{\E} = \sum E^i g_{ij} E^j = \Lambda_l,\,
{\M}\cdot{\M} = \sum M^i g_{ij} M^j = \Lambda_m,\,
{\S}\cdot{\S} = \sum S^i g_{ij} S^j = \Lambda_s.$

In Lagrangian space, the metric tensor $g_0^{ij}$ of the Lagrangian
coordinates (which are taken to be Cartesian coordinates) is the 
unit matrix.
Hence ${\e} ({\m}, {\s})$ can not be distinguished from ${\E} ({\M}, {\S})$
and one can label them with ${\e}_0,\,{\m}_0$ and ${\s}_0$ for clarity.
It is easy to see that ${\e}_0\cdot{\e}_0=\sum e_i g_0^{ij} e_j =1,\,
{\m}_0\cdot{\m}_0 = {\s}_0\cdot{\s}_0 = 1,$ and
${\e}_0\cdot{\m}_0 = {\e}_0\cdot{\s}_0 = {\m}_0\cdot{\s}_0=0.$ 
Most discussions in the paper are within Lagrangian space, so we drop
the subscript for simplicity. Hence ${\s}$ in the main body of the paper
should be understood as ${\s}_0$ and ${\s}_\infty$ is the time asymptotic limit
of ${\s}_0.$

\section{The derivatives of the metric tensor}
\label{appendix:derivative}
The spatial derivative of the metric tensor in Lagrangian coordinates
is
\begin{eqnarray}
{{\partial g_{ij}}\over{\partial \xi^k}} 
= & &
{{\partial \Lambda_l}\over{\partial \xi^k}} e_i e_j
+ \Lambda_l {\partial e_i\over{\partial\xi^k}} e_j
+ \Lambda_l e_i {\partial e_j\over{\partial\xi^k}}   \nonumber \\
& + & {\partial\Lambda_m\over{\partial\xi^k}} m_i m_j
+ \Lambda_m {\partial m_i\over{\partial\xi^k}} m_j
+ \Lambda_m m_i {\partial m_j\over{\partial \xi^k}}  \nonumber \\
& + & {\partial\Lambda_s\over{\partial\xi^k}} s_i s_j
+ \Lambda_s {\partial s_i\over{\partial \xi^k}} s_j
+ \Lambda_s s_i {\partial s_j\over{\partial \xi^k}}.
\end{eqnarray}
Using the various orthonormal relationships outlined in 
appendix \ref{appendix:metric}, one finds
\begin{eqnarray}
{\S}\cdot{\partial\tensor{\bf g}\over{\partial\xi^k}}\cdot{\S}
& = & 
{\partial\Lambda_s\over{\partial\xi^k}}; \\
{\E}\cdot{\partial\tensor{\bf g}\over{\partial\xi^k}}\cdot{\S}
& = &
(\Lambda_l - \Lambda_s) {\e}_0\cdot {\partial {\s}_0\over{\partial\xi^k}}; \\ 
{\M}\cdot{\partial\tensor{\bf g}\over{\partial\xi^k}}\cdot{\S}
& = &
(\Lambda_m - \Lambda_s) {\m}_0\cdot {\partial {\s}_0\over{\partial\xi^k}},
\end{eqnarray}
where ${\e}_0, {\m}_0,$ and ${\s}_0$ are orthonormal vectors in Lagrangian space,
appendix \ref{appendix:metric}.
The spatial derivative of vector ${\s}_0$ is given by
\begin{equation}
{\partial{\s}_0\over{\partial\xi^k}} =
[{\E}\cdot{\partial\tensor{\bf g}\over{\partial\xi^k}}\cdot{\S} / 
(\Lambda_l - \Lambda_s)] {\e}_0
+ [{\M}\cdot{\partial\tensor{\bf g}\over{\partial\xi^k}}\cdot{\S} /
(\Lambda_m - \Lambda_s)] {\m}_0.
\end{equation}
The divergence of ${\s}_0$ vector can be found from the various component of
this equation.
Since $\Lambda_s = \exp (2\lambda_s t) = \exp (-2\lambda t),$
the spatial derivative of the finite time Lyapunov exponent is 
related to the derivatives of the metric tensor by
\begin{equation}
{\partial \lambda t \over{\partial\xi^k}} = - {1\over{2\Lambda_s}}
\Bigl({\S}\cdot{\partial\tensor{\bf g}\over{\partial\xi^k}}\cdot{\S}\Bigr)
\end{equation}
Hence ${\s}_0\cdot\nabla_0 \lambda t + \nabla_0\cdot{\s}_0$ can be 
directly calculated using the spatial derivatives of the metric tensor,
which have analytical expressions if the flow field is specified
in the form of an explicit function of space and time.
 
\section{Convergence of ${\e}, {\m}$ and ${\s}$ vectors in an integrable region of the flow}
\label{appendix:convergence}

If we write the eigenvectors of the
 metric tensor $\tensor{\bf g}$ (Eq.\,(\ref{metric})) of the Lagrangian
coordinates $\Psi_0$-$\varphi_0$-$\vartheta_0$ 
 in their covariant components, {\it i.e.},
${\e} = (e_\Psi, e_\varphi, e_\vartheta),\, {\m} = (m_\Psi, m_\varphi,m_\vartheta)$
and ${\s} = (s_\Psi, s_\varphi, s_\vartheta),$ one finds
 that in an integrable region of the flow, 
\begin{eqnarray}
{e_\Psi\over e_\vartheta} & = & {{{\cal F}{\cal A}^2 + {\cal H}{\cal B}^2 +
         2{\cal G}{\cal A}{\cal B}}\over
        {{\cal G}{\cal A} + {\cal H}{\cal B}}} t 
        + {{2{\cal G}{\cal D}{\cal A}^2 - {\cal F}{\cal E}{\cal A}^2 +
         {\cal H}{\cal E}{\cal B}^2 + 2{\cal D}{\cal H}{\cal A}{\cal B}}
        \over{({\cal G}{\cal A} + {\cal H}{\cal B})^2}} + {\cal O}(t^{-1}); \\ 
{e_\varphi\over e_{\vartheta}} & = & {{{\cal F}{\cal A} + {\cal G}{\cal B}}\over{
        {\cal G}{\cal A} + {\cal H}{\cal B}}} + {\cal O}(t^{-1});\\
{m_\Psi\over m_\vartheta} & = & {{-{\cal G}{\cal A}^2 - {\cal H}{\cal A}{\cal B}
        + {\cal F}{\cal A}{\cal B} + {\cal G}{\cal B}^2}\over{2{\cal G}{\cal A}^2{\cal B}
        + {\cal H}{\cal A}{\cal B}^2 + {\cal F}{\cal A}^3}}{1\over t} + {\cal O}(t^{-2});\\
{m_\varphi\over m_\vartheta} & = & -{{\cal B}\over{\cal A}} + {\cal O}(t^{-1});\\
{s_\Psi\over s_\vartheta} & = & -{1\over {{\cal B}t}} + {\cal O}(t^{-2});\\
{s_\varphi\over s_\vartheta} & = & {{\cal A}\over {\cal B}} + {\cal O}(t^{-1}).
\end{eqnarray}
Hence in the integrable region of the flow, 
${\e},\,{\m}$ and ${\s}$ vectors converge linearly in time to their 
time asymptotic limits, ${\e}_\infty \propto (1,0,0),\, {\m}_\infty \propto
 (0,{\cal B},-{\cal A})$ and ${\s}_\infty \propto (0,{\cal A},{\cal B}).$

\section{Diffusivity coefficients}
\label{appendix:diffusivity}

\begin{eqnarray}
c_0 & = & ({\cal F}{\cal H}-{\cal G}^{2})/J_0^2; \\ 
c_1 & = & ({\cal B}{\cal H}{\cal D}-{\cal B}{\cal G}{\cal E}+
        {\cal A}{\cal G}{\cal D}-{\cal A}{\cal F}{\cal E})/(J_0^2
        \sqrt{{\cal A}^2 + {\cal B}^2}); \\ 
c_2 & = & (-{\cal A}{\cal H}{\cal D}+
        {\cal A}{\cal G}{\cal E}+{\cal B}{\cal G}{\cal D}-
        {\cal B}{\cal F}{\cal E})/(J_0^2\sqrt{{\cal A}^2 + {\cal B}^2}); \\
c_3 & = & {{{\cal B}^{2}{\cal H}{\cal C}-{\cal B}^{2}{\cal E}^{2}+
        2\,{\cal B}{\cal A}{\cal G}{\cal C}-
        2\,{\cal B}{\cal A}{\cal D}{\cal E}+
        {\cal A}^{2}{\cal F}{\cal C}-{\cal A}^{2}{\cal D}^{2}} 
        \over{J_0^2({\cal A}^2 + {\cal B}^2)}}; \\
c_4 & = & { {{\cal B}^{2}
        {\cal G}{\cal C}-{\cal A}^{2}{\cal G}{\cal C}+
        {\cal A}^{2}{\cal D}{\cal E}-
        {\cal B}{\cal A}{\cal H}{\cal C}-
        {\cal B}^{2}{\cal D}{\cal E}-{\cal A}{\cal B}{\cal D}^{2}+
        {\cal B}{\cal A}{\cal E}^{2}+
        {\cal A}{\cal B}{\cal F}{\cal C}}
        \over{J_0^2({\cal A}^2 + {\cal B}^2)}}; \\
c_5 & = & {{{\cal B}^{2}{\cal F}{\cal C}-{\cal B}^{2}{\cal D}^{2}+
        {\cal A}^{2}{\cal H}{\cal C}+2\,{\cal B}{\cal A}{\cal D}{\cal E}-
        2\,{\cal B}{\cal A}{\cal G}{\cal C}-
        {\cal A}^{2}{\cal E}^{2}}\over
        {J_0^2({\cal A}^2 + {\cal B}^2)}};
\end{eqnarray}
 
Here $J_0^2$ is the determinant of the metric tensor of the $\Psi$-$\varphi$-$\vartheta$
coordinates, Eq. (\ref{s4e7}).

\newpage

\centerline{Figures}

FIG. 1: The finite time Lyapunov exponent
$\lambda$ is related to the  largest eigenvalue of the metric tensor $\protect{\Lambda}$
by $\ln\protect{\Lambda} = 2\lambda t$ (uptriangles).
$\theta$ and $\phi$ are the polar and azimuthal angles of the ${\s}$ vector,
$\ln(d\theta/dt)$ (downtriangles) and $\ln(d\phi/dt)$ (circles).
($a$) Standard map with $k=1.5,$ at point
$(0.3,0.6);$ ($b$) Extended standard map with $k=1.5$ 
and $\Delta=\sqrt{3},$ 
at point
$(0.3,0.6,0.8)$. \\

FIG. 2: The distribution of finite time Lyapunov
exponents along a single trajectory peaks around the infinite time Lyapunov
exponent.The finite time Lyapunov exponents, $\lambda(\xi,t),$ are
evaluated at fixed $t.$ 
Extended standard map with $k=3.0$ and
$\Delta=\sqrt{3},\, t=20$ iterations. \\

FIG. 3: The residue, or difference, between the distribution of 
finite time Lyapunov exponents
 and a Gaussian distribution,
decreases as $t$ increases. Circles are for extended standard map with $k=10.0$
and $\Delta^2=3.$ Triangles are for standard map with $k=10.0.$
Dashed and Solid lines are given by $Residue=0.347/\sqrt{t} + 0.018$ and
 $Residue=0.31/\sqrt{t}+0.0026.$ \\

FIG. 4: The standard deviation of the distribution of 
finite time Lyapunov exponents
 decreases the further the flow is  from being integrable (larger $k$).
Uptriangles and circles are for extended standard map with $t=40$ and $20$ iterations,
respectively. Downtriangles are for standard map with $t=40$ iterations.\\

FIG. 5: The standard deviation of the
 distribution of finite time Lyapunov exponents
 scales as $1/\sqrt{t}.$ Uptriangles are for standard map with $k=10.0.$
Circles are for extended standard map with $k=10.0$ and $\Delta=\sqrt{3}.$
Dashed and Solid lines are given by $\sigma=0.785/\sqrt{t}$ and 
$\sigma=0.382/\sqrt{t}.$ \\

FIG. 6: $\Delta(\xi,t)\equiv |{\s}\cdot\nabla_0(\lambda t) 
+ \nabla_0\cdot{\s}|$
exponentially converges to zero. Uptriangle: $\Delta(\xi,t)$ is evaluated at
point $(0.3,0.6,0.8)$ for extended standard map with $k=1.5.$
Circles: $\Delta(\xi,t)$ is evaluated at point $(0.1,0.2,0.75)$ for ABC map with 
$A=B=C=1.$ \\

FIG. 7: Extended standard map with $k=2.0$ and $\Delta=\sqrt{3}.$
 ($a$) The Lyapunov exponents,\, $\lambda(t=30),$ were sampled with equal spacing 
along an ${\e}$ line, an ${\m}$ line and an ${\s}$ line 
(all starting at (0.1,0.1,0.8)). These values are plotted
against the distance along the lines;
($b$) The correlation function for the Lyapunov exponents in
 the ${\s}$ direction (solid),  in the ${\m}$ direction (dashed)
and in the ${\e}$ direction (dotted). \\

FIG. 8: The finite time Lyapunov exponent ($\lambda$), the ${\e}$ and ${\m}$ components
of the ${\s}$ line curvature ($\kappa_e$ and $\kappa_m$), 
are plotted as functions of distance along an ${\s}$ line.
The calculation was done for ABC map with $A=B=C=1.$
Only the magnitudes of the curvature are used for the log-linear plots. 


\begin{references}
\bibitem{comment_lagrangian} We note that previous analyses by 
        others \cite{ottino2,wiggins} have employed a Lagrangian approach, but they are 
        only locally valid, in contrast to the global (Lagrangian) coordinate
        transformation used in \cite{tang} and this paper.
\end{references}
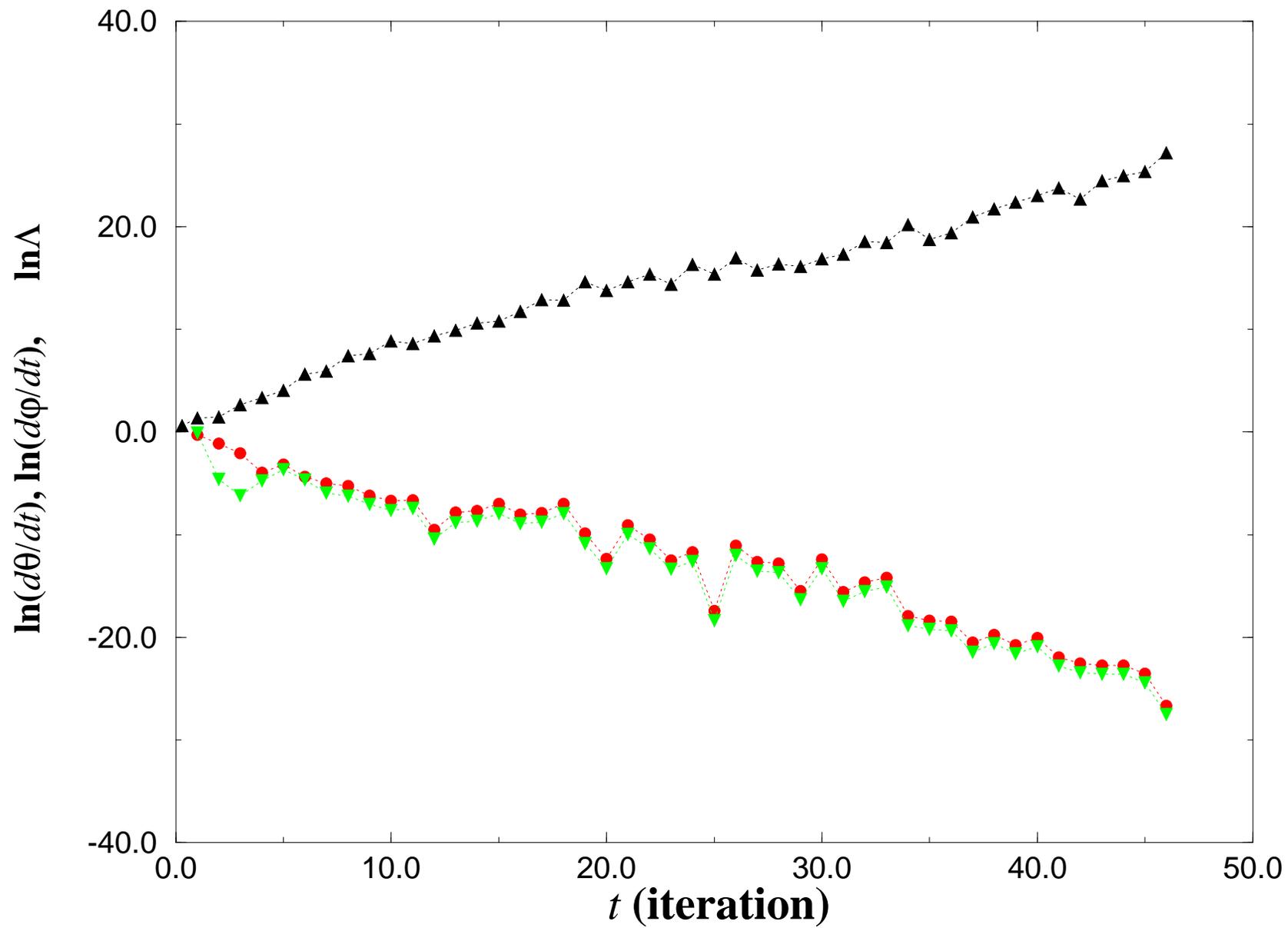
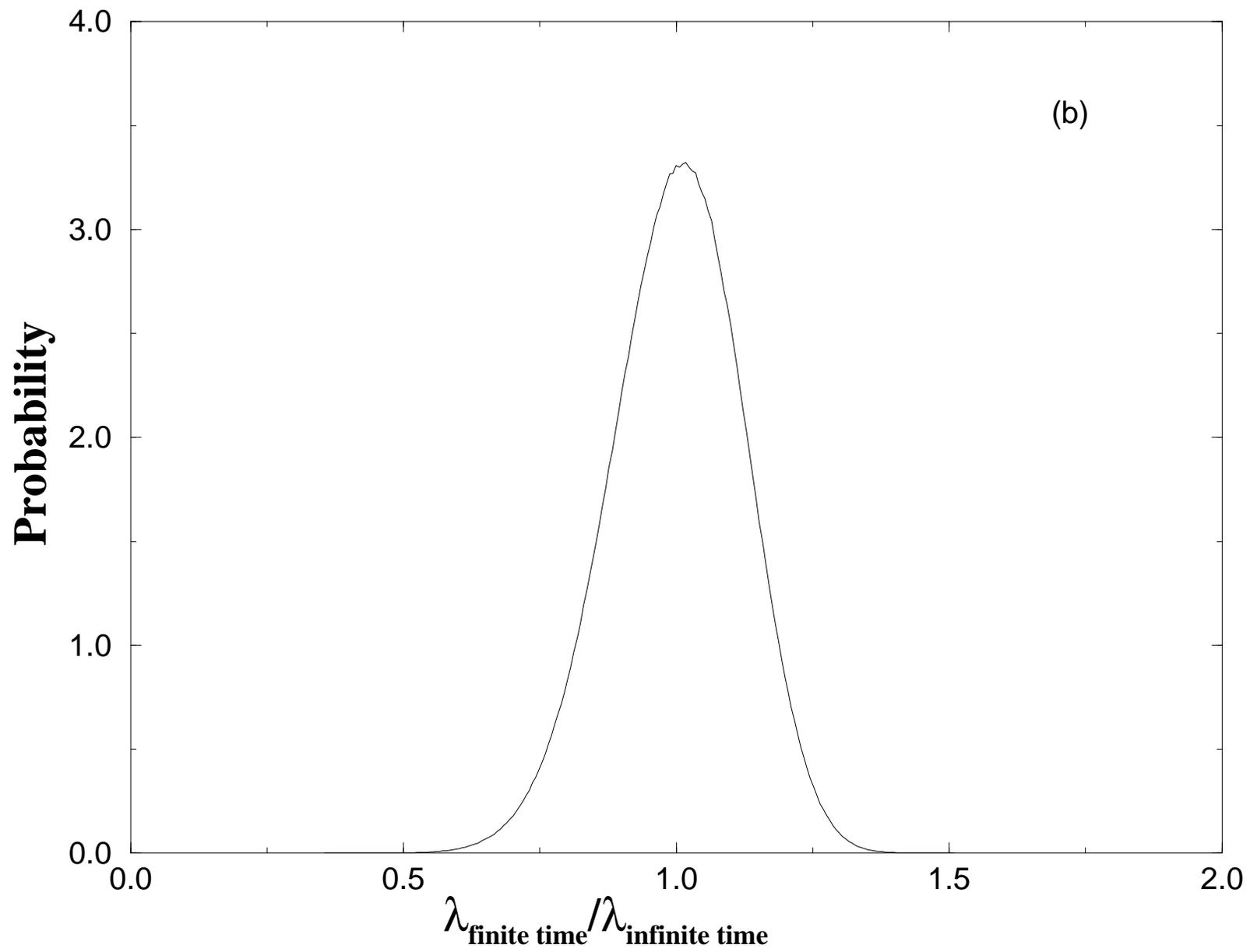
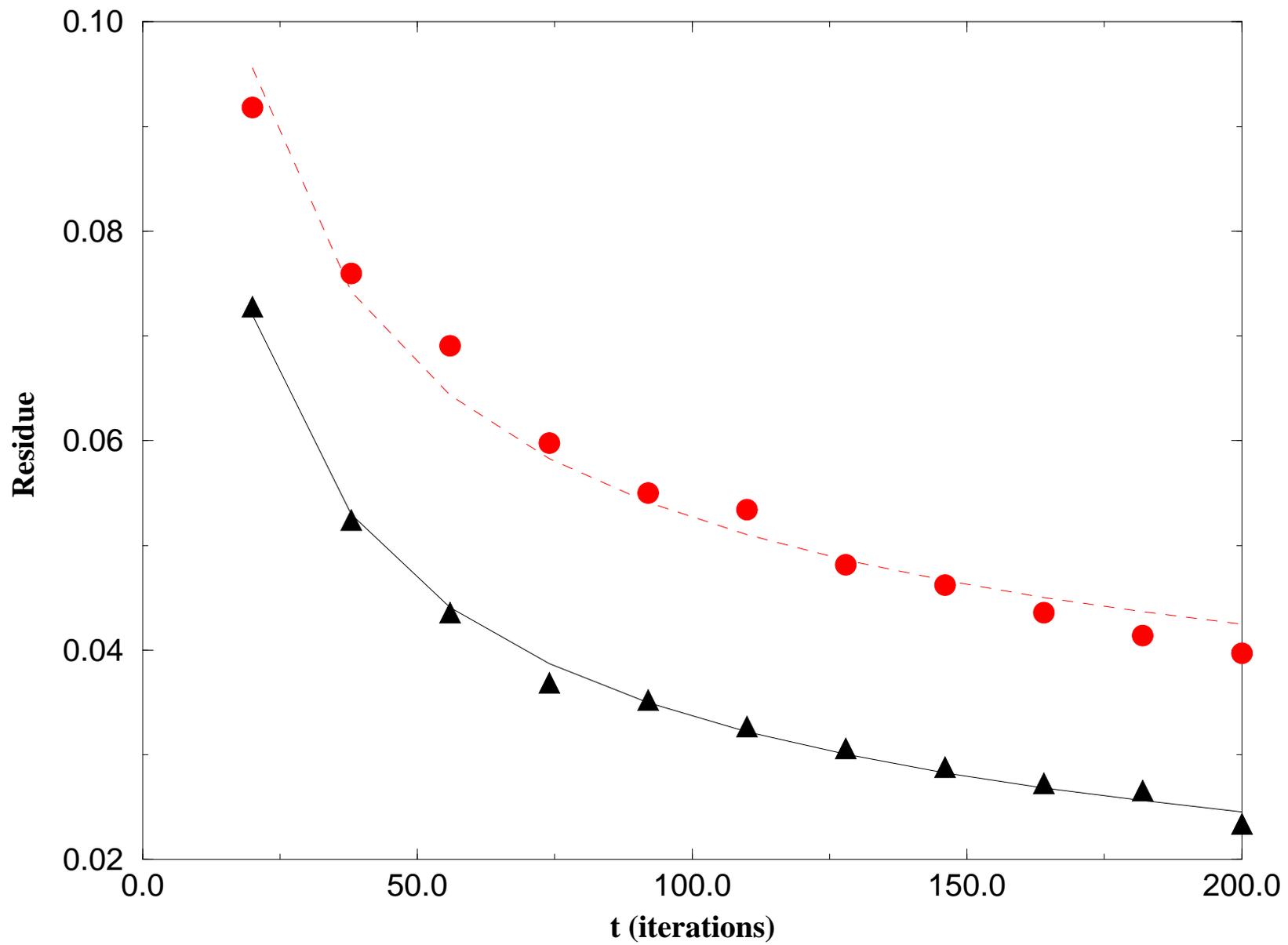
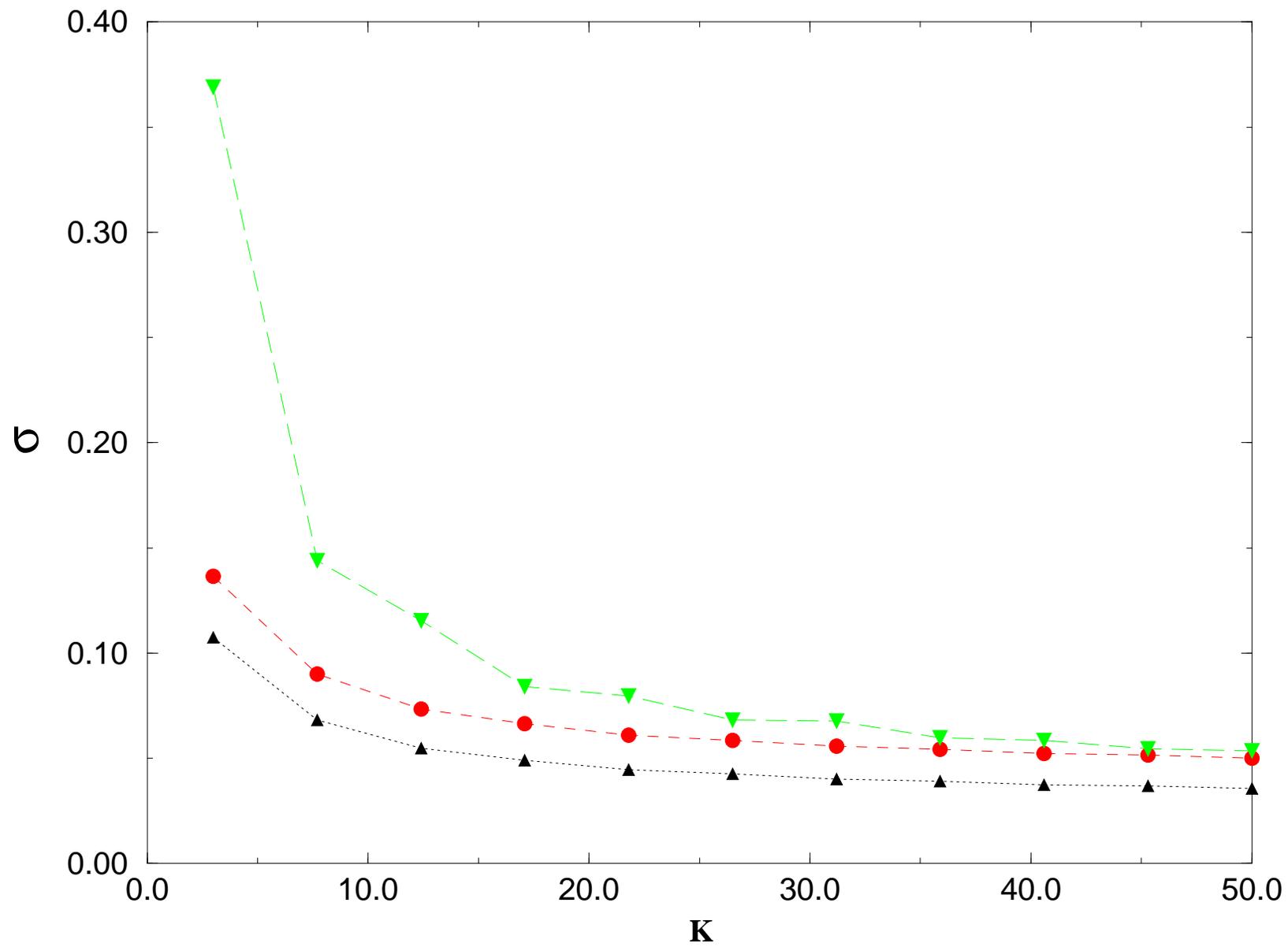
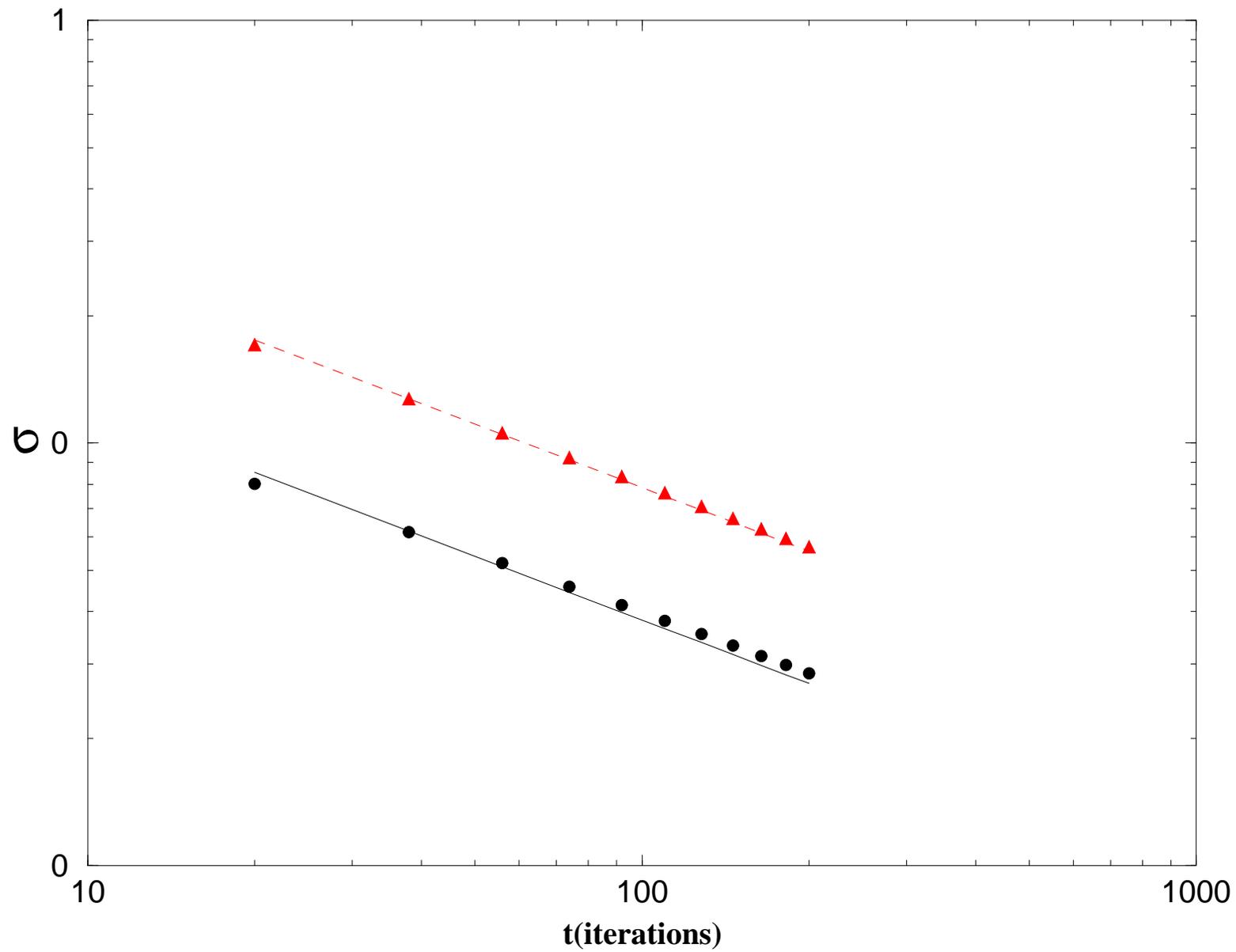
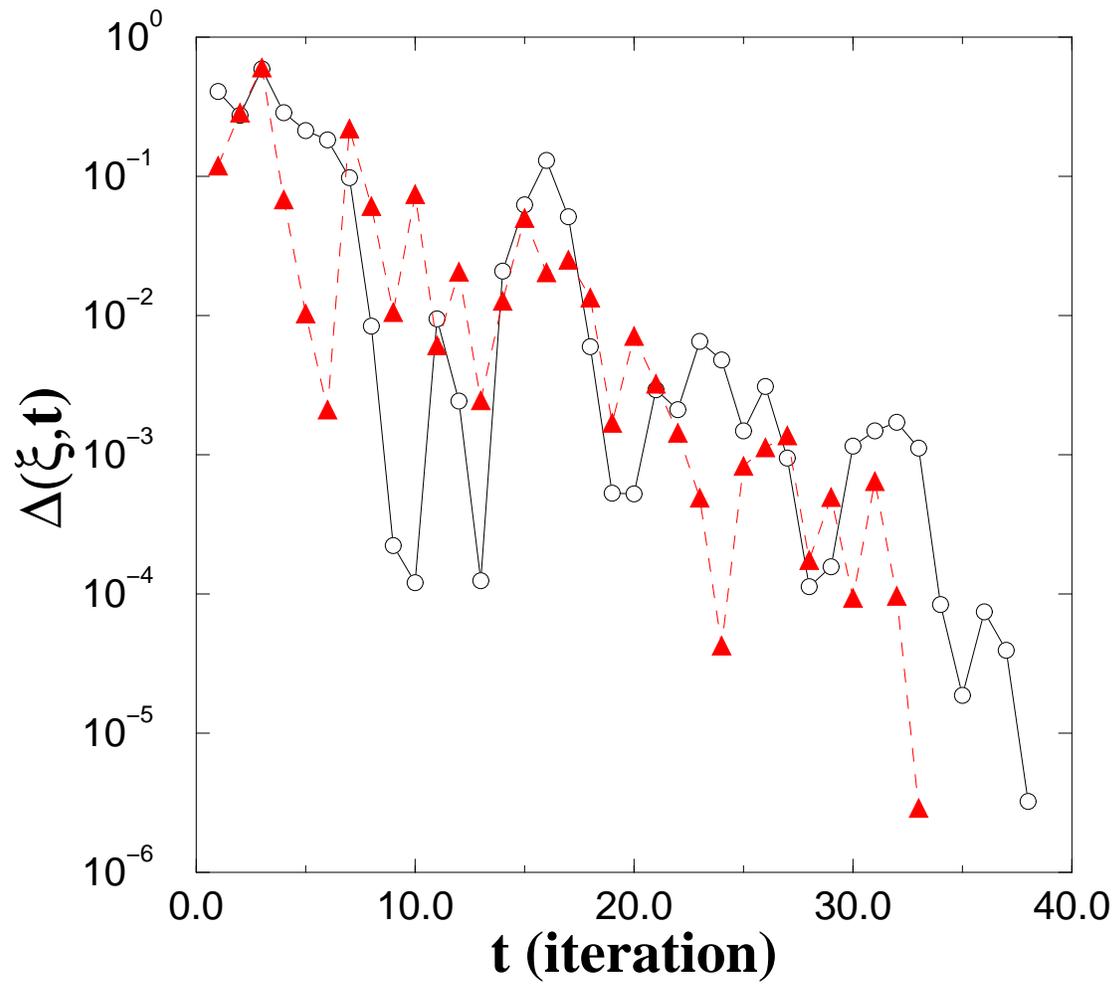
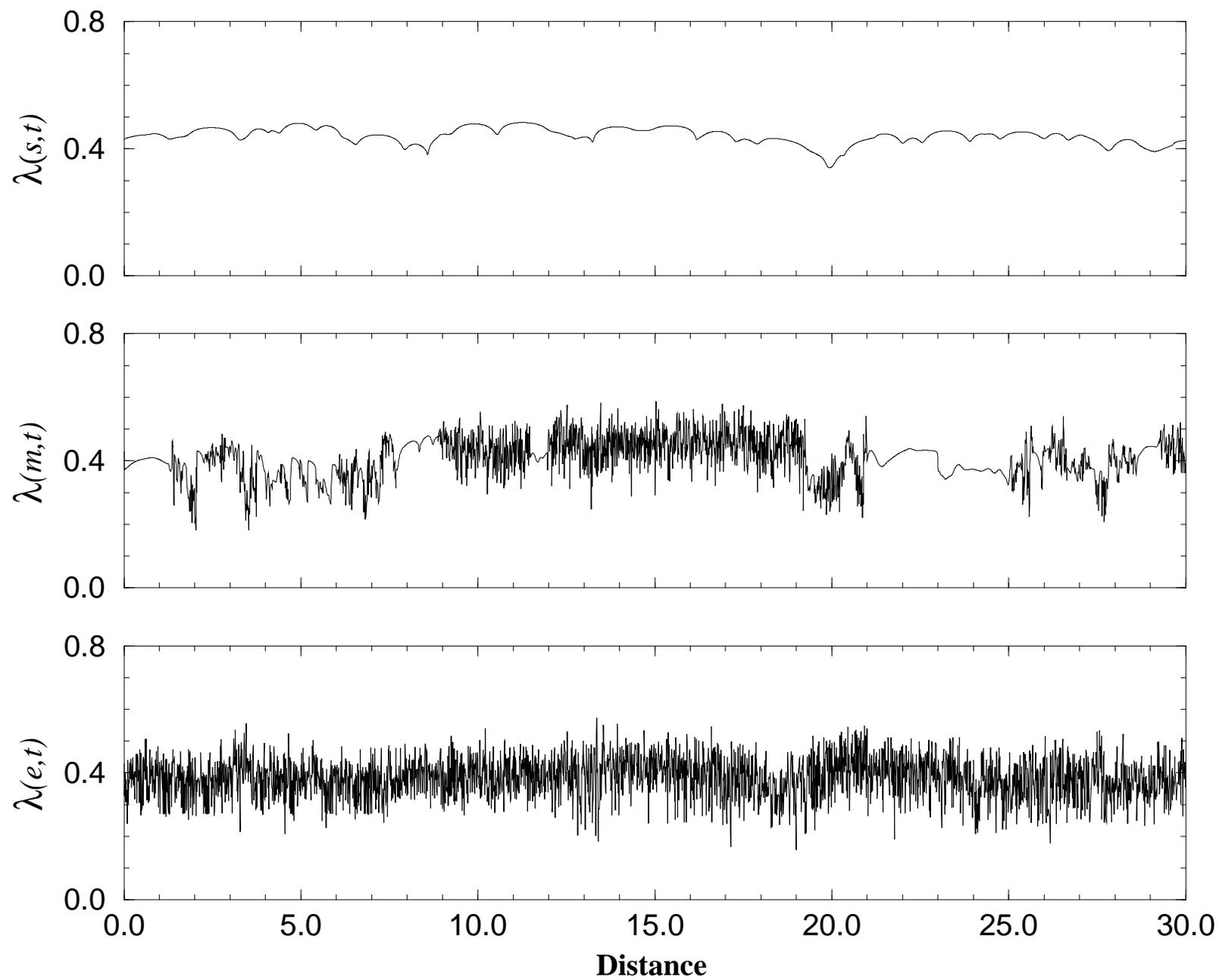
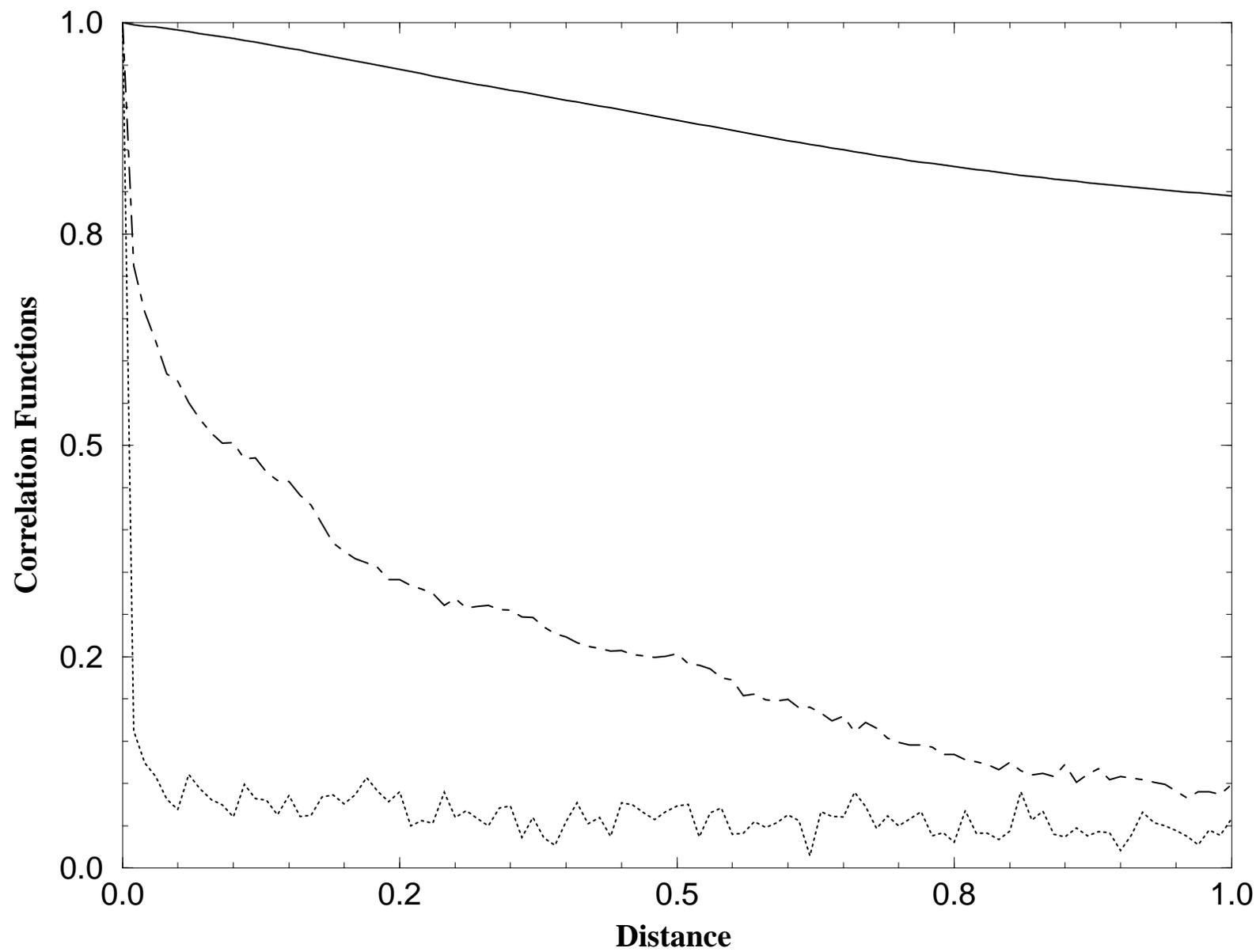
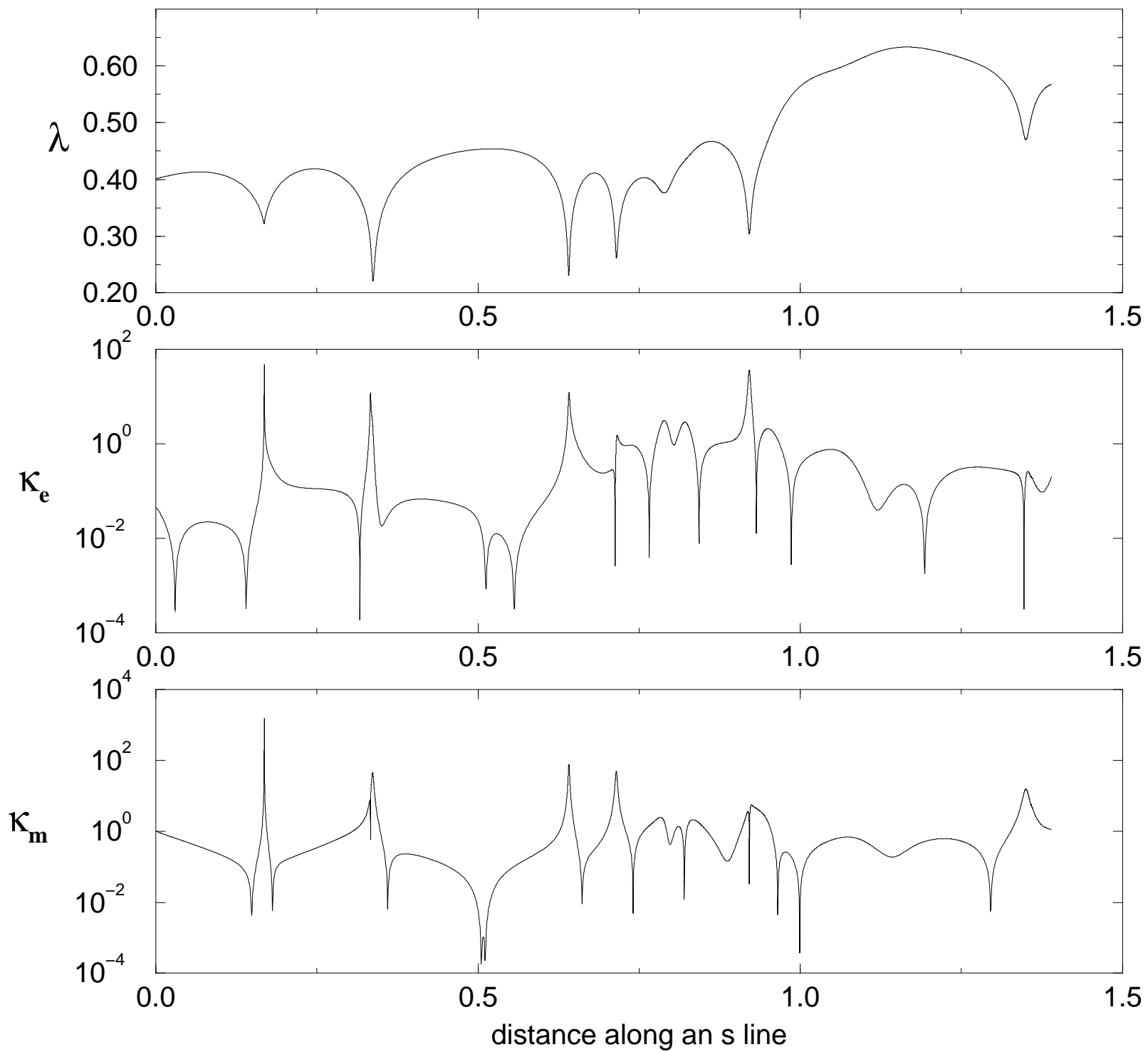
\end{document}